\documentclass [aip, jmp, reprint, superscriptaddress, eqsecnum, onecolumn]{revtex4-1}
\usepackage {etex}

\usepackage [english]{babel}
\usepackage [T1]{fontenc}
\usepackage [utf8] {inputenc}
\usepackage {lmodern}
\usepackage {geometry}
\usepackage {amsfonts}
\usepackage {amsmath}	
\usepackage {amssymb}
\usepackage {amsxtra}
\usepackage [amsthm,thmmarks,amsmath]{ntheorem} 
\usepackage {wasysym}
\usepackage {mathtools}
\usepackage {calrsfs}
\usepackage {dsfont}
\usepackage {graphicx}
\usepackage {rotating}

\usepackage {setspace}
\usepackage {color}
\usepackage [all,cmtip]{xy}
\usepackage {hyperref}
\usepackage [nottoc,numbib]{tocbibind}
\usepackage {diagbox}
\usepackage {dcolumn}

\allowdisplaybreaks

\input diagxy

\DeclareFontFamily{OT1}{pzc}{}
\DeclareFontShape{OT1}{pzc}{m}{it}{<-> s * [1.30] pzcmi7t}{}
\DeclareMathAlphabet{\mathpzc}{OT1}{pzc}{m}{it}

\DeclareMathOperator{\R}{\mathbb R}

\DeclareMathOperator{\C}{\mathbb C}
\DeclareMathOperator{\N}{\mathbb N}

\DeclareMathOperator{\Cyl}{\textup{Cyl}}

\DeclareMathOperator{\uR}{\textup R}
\DeclareMathOperator{\uL}{\textup L}

\DeclareMathOperator{\uP}{\textup P}

\DeclareMathOperator{\fg}{\mathfrak g}

\DeclareMathOperator{\fA}{\mathfrak A}
\DeclareMathOperator{\fB}{\mathfrak B}
\DeclareMathOperator{\fC}{\mathfrak C}

\DeclareMathOperator{\fH}{\mathfrak H}

\DeclareMathOperator{\fS}{\mathfrak S}
\DeclareMathOperator{\fX}{\mathfrak X}
\DeclareMathOperator{\cA}{\mathcal A}

\DeclareMathOperator{\cG}{\mathcal G}

\DeclareMathOperator{\cD}{\mathcal D}

\DeclareMathOperator{\cL}{\mathcal L}

\DeclareMathOperator{\aut}{\textup{Aut}}

\DeclareMathOperator{\hol}{\textup hol}

\DeclareMathOperator{\id}{id}

\newtheoremstyle{breakdef}%
  {\item[\rlap{\vbox{\normalfont\bfseries\hbox{\llap{##2}\hskip\labelsep
          ##1:}\hbox{\\[0.1cm]}}}]}%
  {\item[\rlap{\vbox{\normalfont\bfseries\hbox{\llap{##2}\hskip\labelsep
          ##1 (##3):}\hbox{\\[0.1cm]}}}]}
\newtheoremstyle{breaksatz}%
  {\item[\rlap{\vbox{\normalfont\normalsize\bfseries\hbox{\llap{##2}\hskip\labelsep
          ##1:}\hbox{\\[0.1cm]}}}]}%
  {\item[\rlap{\vbox{\normalfont\normalsize\bfseries\hbox{\llap{##2}\hskip\labelsep
          ##1 (##3):}\hbox{\\[0.1cm]}}}]}
\newtheoremstyle{breaklem}%
  {\item[\rlap{\vbox{\normalfont\normalsize\bfseries\hbox{\llap{##2}\hskip\labelsep
          ##1:}\hbox{\\[0.1cm]}}}]}%
  {\item[\rlap{\vbox{\normalfont\normalsize\bfseries\hbox{\llap{##2}\hskip\labelsep
          ##1 (##3):}\hbox{\\[0.1cm]}}}]}
\newtheoremstyle{breakprop}%
  {\item[\rlap{\vbox{\normalfont\normalsize\bfseries\hbox{\llap{##2}\hskip\labelsep
          ##1:}\hbox{\\[0.1cm]}}}]}%
  {\item[\rlap{\vbox{\normalfont\normalsize\bfseries\hbox{\llap{##2}\hskip\labelsep
          ##1 (##3):}\hbox{\\[0.1cm]}}}]}
\newtheoremstyle{breakbem}%
  {\item[\rlap{\vbox{\hbox{\hskip\labelsep\normalfont\bfseries
          ##1 ##2:}\hbox{\\[0.1cm]}}}]}%
  {\item[\rlap{\vbox{\hbox{\hskip\labelsep\normalfont\bfseries
          ##1 ##2 (##3):}\hbox{\\[0.1cm]}}}]}
\newtheoremstyle{breakbsp}%
  {\item[\rlap{\vbox{\hbox{\hskip\labelsep\normalfont\bfseries
          ##1 ##2:}\hbox{\\[0.2cm]}}}]}%
  {\item[\rlap{\vbox{\hbox{\hskip\labelsep\normalfont\bfseries
          ##1 ##2 (##3):}\hbox{\\[0.2cm]}}}]}
\newtheoremstyle{breakkor}%
  {\item[\rlap{\vbox{\hbox{\hskip\labelsep\normalfont\bfseries
          ##1 ##2:}\hbox{\\[0.1cm]}}}]}%
  {\item[\rlap{\vbox{\hbox{\hskip\labelsep\normalfont\bfseries
          ##1 ##2 (##3):}\hbox{\\[0.1cm]}}}]}
\newtheoremstyle{proof}%
  {\item[\rlap{\vbox{\hbox{\hskip\labelsep\normalfont\bfseries
          \underline{##1:}}\hbox{\\[0.1cm]}}}]}%
  {\item[\rlap{\vbox{\hbox{\hskip\labelsep\normalfont\bfseries
          \underline{##1 (##3):}}\hbox{\\[0.1cm]}}}]}
\theoremstyle{breakkor} 
\newtheorem{Definition}{Definition}[section] 
\theoremstyle{breakkor}
\newtheorem{Theorem}[Definition]{Theorem}

\theoremstyle{breakkor}
\newtheorem{Lemma}[Definition]{Lemma}
\theoremstyle{breakkor}
\newtheorem{Proposition}[Definition]{Proposition}
\theoremstyle{breakkor}
\newtheorem{Corollary}[Definition]{Corollary}
\theoremstyle{breakkor}
\theorembodyfont{\normalfont}
\newtheorem{Remark}[Definition]{Remark}
\newtheorem{Commentary}[Definition]{Commentary}

\theoremstyle{proof}
\theoremsymbol{\fbox{}}
\newtheorem{Proof}{Proof}

\begin{document}
\title{Coherent states, quantum gravity and the Born-Oppenheimer approximation, III: Applications to loop quantum gravity}
\author{Alexander Stottmeister}
\email{alexander.stottmeister@gravity.fau.de}
\author{Thomas Thiemann}
\email{thomas.thiemann@gravity.fau.de}
\affiliation{Institut für Quantengravitation, Lehrstuhl für Theoretische Physik III, Friedrich-Alexander-Universität Erlangen-Nürnberg, Staudtstraße 7/B2, D-91058 Erlangen, Germany}
\begin{abstract}
In this article, the third of three, we analyse how the Weyl quantisation for compact Lie groups presented in the second article of this series fits with the projective-phase space structure of loop quantum gravity-type models. Thus, the proposed Weyl quantisation may serve as the main mathematical tool to implement the program of space adiabatic perturbation theory in such models. As we already argued in our first article, space adiabatic perturbation theory offers an ideal framework to overcome the obstacles that hinder the direct implementation of the conventional Born-Oppenheimer approach in the canonical formulation of loop quantum gravity. 
\end{abstract}
\maketitle
\tableofcontents
\section{Introduction}
\label{sec:intro} 
In our previous articles in this series\cite{StottmeisterCoherentStatesQuantumI, StottmeisterCoherentStatesQuantumII}, we pointed out the need for a Weyl quantisation for models of loop quantum gravity-type to realise the (time-dependent) Born-Oppenheimer approximation for multi-scale quantum dynamical systems along the lines of space adiabatic perturbation theory\cite{PanatiSpaceAdiabaticPerturbation}. In the second article of this series, we introduced a Weyl quantisation for compact Lie groups and developed the basis for an associated calculus of Paley-Wiener-Schwartz symbols, which allowed us to tackle the ``problem of  non-commutative fast-slow coupling'' (originally pointed out in the context of loop quantum gravity\cite{GieselBornOppenheimerDecomposition}).\\
But, if we intend to use the Born-Oppenheimer approach to extract a continuum limes in the slow (gravitational) sector (cf. \cite{GieselBornOppenheimerDecomposition, StottmeisterCoherentStatesQuantumI}), there is a second obstacle. The latter can be addressed in terms of compatibility conditions of the Weyl quantisation with the projective limit structures involved in the construction of the models {\`a} la loop quantum gravity. We expect, that such compatibility conditions, in addition to a selection of admissible observables, play a major role in the possible extraction of quantum field theory on curved spacetimes from loop quantum gravity (with matter). \\
This said, it is the primary objective of the present article to investigate the possibility of formulating a Weyl quantisation suitable for loop quantum gravity-type models. \\[0.1cm]
Before we come to the main part of the article, let us briefly outline its structure and content:\\[0.1cm]
The main section \ref{sec:apploopphase} is devoted to applications of the (abstract) methods introduced in the previous article\cite{StottmeisterCoherentStatesQuantumII}. In the first subsection \ref{subsec:projectweyl}, we apply the global and local Weyl quantisations for compact Lie groups to the basic building blocks of loop quantum gravity-type models, $T^{*}G$, $G$ a compact Lie group. Moreover, we show how and to what extent compatibility with the projective limit, $\overline{\Gamma}=\varprojlim_{l\in\cL}\Gamma_{l},\ \Gamma_{l}\cong T^{*}G^{n_{l}}$ (cf. \cite{ThiemannQuantumSpinDynamics7}), of finite dimensional truncations of the gravitational phase space, $\Gamma=|\Lambda|^{1}T^{*}\mathcal{A}_{\uP}$ (in Ashtekar-Barbero variables), can be achieved, thus, allowing for a genuine Weyl quantisation of loop quantum gravity-type models. In the course of this analysis, we discover certain subtle differences between the phase space quantisation and the quantisation in terms of the holonomy-flux algebra, that was so far only noticed in recent work by Lanéry and Thiemann\cite{LaneryProjectiveLoopQuantum}. In respect of the Born-Oppenheimer approximation, the main difference of our approach to previous ones (notably \cite{GieselBornOppenheimerDecomposition}) is that we aim, already from the beginning, for a technical setup, which is able to deal with full loop quantum gravity (in its common realisations).\\
In subsection \ref{subsec:indlim}, we analyse the possibility to define a ``non-commutative phase space’’ by means of the inductive family of quantum algebras that is obtained from the Weyl quantisation of the projective family of (truncated) phase spaces. We also comment on the dual notion of projective families of (algebraic) state spaces (cp. \cite{LaneryProjectiveLoopQuantum}).\\
In the last subsection \ref{subsec:gauge} of the main part, we explain the behaviour of gauge transformation w.r.t. the Weyl quantisation and the projective/inductive limit structures.\\
Finally, we present some concluding remarks and perspectives in section \ref{sec:con}.
\section{Loop quantum gravity and phase space quantisation}
\label{sec:apploopphase}
While the previous articles\cite{StottmeisterCoherentStatesQuantumI,StottmeisterCoherentStatesQuantumII} were of a rather general mathematical character, the present section is devoted to discussing applications of the outlined framework to models of a loop quantum gravity-type (cf. \cite{StottmeisterCoherentStatesQuantumI} for an application to spin systems). We show how the transformation group $C^{*}$-algebra $C(G)\rtimes_{\uL}G$ makes a, quite natural, appearance in the phase space quantisation of loop quantum gravity type models that are based on a gauge theory with compact (Lie) structure group $G$, and discretisations w.r.t. graphs (cf. \cite{ThiemannQuantumSpinDynamics7} and references therein). Furthermore, we discuss how Weyl and Kohn-Nirenberg quantisation enter the picture. To begin with, we recall some basic notions from loop quantum gravity. We follow closely \cite{ThiemannQuantumSpinDynamics7, StottmeisterStructuralAspectsOf}, although we refine certain aspects of the presentation:\\[0.1cm]
Loop quantum gravity is based on a Hamiltonian formulation of general relativity in terms of a constrained Yang-Mills-type theory, i.e. in a field theoretic description the phase space of the classical theory is given by the (densitiesed) cotangent bundle $|\Lambda|^{1}T^{*}\mathcal{A}_{\uP}$ to the space of connections $\cA_{\uP}$ on a given (right, semi-analytic\footnote{An elementary introduction to the semi-analytic category can be found in \cite{ThiemannModernCanonicalQuantum}.}) principal $G$-bundle $\uP\stackrel{\pi}{\rightarrow}\Sigma$, where $\Sigma$ is the spatial manifold in a 3+1-splitting of a (globally hyperbolic) spacetime $\textup{M}\cong\R\times\Sigma$. In general relativity, we have $G=\textup{SU(2)}, \textup{Spin}_{4}$, or central quotients of these groups, but for most of what follows we only need to assume that $G$ is a compact Lie group.\\[0.1cm]
The basic variables, the theory is phrased in, are the Ashtekar-Barbero connection $A\in\mathcal{A}_{\uP}$ and its conjugate momentum $E\in\Gamma\left(T\Sigma\otimes\textup{Ad}^{*}(\uP)\otimes|\Lambda|^{1}(\Sigma)\right)$. Strictly speaking, we further require $E$ to be non-degenerate as a (densitiesed) section of the bundle of linear operators $\textup{L}(\textup{Ad}(\uP),T\Sigma)$. In general relativity, the existence of $E$ is ensured by the triviality of the orthogonal frame bundle $P_{\textup{SO}}(\Sigma)$. This mathematical setup also appears to be valid in the context of the new variables proposed in \cite{BodendorferNewVariablesFor1, BodendorferNewVariablesFor2}. Here, $\textup{Ad}^{*}(\uP):=\uP\times_{\textup{Ad}^{*}}\fg^{*}$ and $|\Lambda|^{1}(\Sigma)$ denotes the bundle of 1-densities on $\Sigma$. Since $\mathcal{A}_{\uP}$ is an affine space modelled on $\Omega^{1}(\textup{Ad}(\uP)):=\Gamma(T^{*}\Sigma\otimes\textup{Ad}(\uP))$, $\textup{Ad}(\uP):=\uP\times_{\textup{Ad}}\fg$ , the following Poisson structure
\begin{align}
\label{eq:poisson}
\{E^{a}_{i}(x),A^{j}_{b}(y)\}=\delta^{a}_{b}\delta^{j}_{i}\delta(x,y)
\end{align}
is meaningful in local coordinates $\phi:U\subset\Sigma\rightarrow V\subset\R^{3}$ subordinate to a local trivialisation $\psi:\uP_{|U}\rightarrow U\times G$, i.e.
\begin{align}
\label{eq:loccoord}
((\phi\circ\psi)^{-1})^{*}A_{|P_{|U}} & =A^{j}_{b}dx^{b}\otimes\tau_{j},\\[0.25cm] \nonumber
(\phi\circ\psi)_{*}E_{|P_{|U}} & =E^{a}_{i}\frac{\partial}{\partial x^{a}}\otimes\tau^{*i}.
\end{align}
Here, $\{\tau_{j}\}_{j}$ is a basis of $\fg$ and $\{\tau^{*i}\}_{i}$ its dual in $\fg^{*}$.\\[0.1cm]
The variables $(A,E)$ are directly related to the Arnowitt-Deser-Misner variables $(q,P)$. Namely, $E^{a}_{i}$ is a densitiesed triad for the spatial metric $q_{ab}E^{a}_{i}E^{b}_{j}=\det(q)\delta_{ij}$, and $A^{i}_{a}=\Gamma^{i}_{a}+K^{i}_{a}$ is built out of the Levi-Civita connection $\Gamma$ of the spatial metric $q$ and the extrinsic curvature $K$ determined by the momentum $P$.\\[0.1cm]
What makes the variables $(A,E)$ special, is that they allow to carry out a canonical quantisation of general relativity, i.e. loop quantum gravity (cf. \cite{RovelliQuantumGravity, ThiemannModernCanonicalQuantum} for general accounts on the topic). Especially, it is possible to construct mathematically well-defined operators for all constraints acting in a suitable Hilbert space within this approach, most prominently the Wheeler-DeWitt constraint (cf. \cite{ThiemannQuantumSpinDynamics1, ThiemannQuantumSpinDynamics2, ThiemannQuantumSpinDynamics3, ThiemannQuantumSpinDynamics4, ThiemannQuantumSpinDynamics5, ThiemannQuantumSpinDynamics6, ThiemannQuantumSpinDynamics7}).
\subsection{Projective-phase space structure and Weyl quantisation}
\label{subsec:projectweyl}
The canonical quantisation of $\Gamma := |\Lambda|^{1}T^{*}\mathcal{A}_{\uP}$, adapted optimally to our framework, starts from the following functionals of the basic variables $(A,E)$:
\begin{Definition}
\label{def:loopphasespacevariables}
Let $\gamma\in\Gamma^{\textup{sa},\uparrow}_{0}$ be an oriented, embedded, semi-analytic, compactly supported, finite graph\index{graph!embedded, semi-analytic, compactly supported, finite!oriented|textbf} in $\Sigma$, $P_{\gamma}$ an  oriented, semi-analytic, polyhedronal decomposition of $\Sigma$ dual to $\gamma$, and $\Pi_{\gamma}$ a system of oriented, semi-analytic paths adapted to $\gamma$ and $P_{\gamma}$, i.e. for every edge $e\in E(\gamma)$ there exists a unique face $S_{e}$ of $P_{\gamma}$ having unique transversal intersection $x_{e}=e\cap S_{e}$ in an interior point, and a collection of paths $\{\rho_{e}(x)\ |\ x\in S_{e}\}\subset\Pi_{\gamma}$ from any $x\in S_{e}$ to $x_{e}$. Moreover, $S_{e}$ carries a compatible orientation of its normal bundle, i.e. aligned with the orientation of the edge $e$. As usual, the edges\index{graph!edge\textbf}\index{edge|see{graph}} (connected, oriented semi-analytic submanifolds, possibly with two-point boundary) and vertices\index{graph!vertex|textbf}\index{vertex|see{graph}} (boundary points of edges) of a graph $\gamma$ are denoted by $e\in E(\gamma)$, respectively, $v\in V(\gamma)$. An edge will be treated as an embedded submanifold $e:[0,1]\rightarrow\Sigma$. The orientations \\[0.1cm]
The triple $l=(\gamma, P_{\gamma}, \Pi_{\gamma})$ is called a \textup{finite, oriented, semi-analytic structured graph}\index{graph!finite, oriented, semi-analytic structured|textbf} in $\Sigma$, and we associate with it the following functionals of $(A,E)\in\Gamma$:\index{functional!holonomy|textbf}\index{functional!momentum/flux|textbf}
\begin{align}
\label{eq:loopphasespacefunctionals}
& g_{e}(A;\sigma)(s,t)\in G,\ s\leq t\in[0,1]\ \textup{s.t.} \hol^{A}_{e}(s(e(s))) = R_{g_{e}(A;\sigma)(s,t)}(s(e(t))), \\[0.25cm] \nonumber
& P^{e}_{X}(A,E;\sigma) = \int_{S_{e}}*\left(\left(\textup{Ad}^{*}_{g_{e}(A;\sigma)(s_{e},1)g_{\rho_{e}}(A;\sigma)(0,1)}(E_{\sigma})\right)(X)\right),\ X\in\fg,
\end{align}
where $\sigma:\Sigma\rightarrow\uP$ is a (measurable) section\index{section!reference|textbf}, which defines a (measurable) equivalences $\Sigma\times G\sim_{\sigma}\uP,\ \psi_{\sigma}(x,g)=R_{g}(\sigma(x)),$ and $\Sigma\times\fg^{*}\sim_{\sigma}\textup{Ad}^{*}(\uP),\ \textup{Ad}^{*}(\psi_{\sigma})(x,\theta) = [(\sigma(x),\theta)]$. In this sense, $E_{\sigma}$ is the vector density valued section of $\Sigma\times\fg$ corresponding to $E$, and $*(E_{\sigma}(X))$ is the pseudo-2-form dual the vector density $E(X)$. The parameter $s_{e}\in [0,1]$ is determined from $e(s_{e})=x_{e}$.\\[0.1cm]
The set of all finite, oriented, semi-analytic structured graphs is called $\cL$, and we abbreviate $g_{e}(A;\sigma)(0,1)=g_{e}(A,\sigma)$ in the following. The images of $\Gamma$ under the functionals \eqref{eq:loopphasespacefunctionals} (s=0,t=1), when varied w.r.t. $l\in\cL$, are the truncated phase spaces $\Gamma_{l}:=\bigtimes_{e\in E(\gamma)}(T^{*}G)_{e}\cong T^{*}G^{|E(\gamma)|}$ (w.r.t. the right trivialisation). The images of $\Gamma$ under the holonomy functionals alone are the truncated configurations spaces $C_{l}:=\bigtimes_{e\in E(\gamma)}G_{e}\cong G^{|E(\gamma)|}$, which are naturally covered by the $\Gamma_{l}$ via the cotangent bundle projection $T^{*}G\rightarrow G$.
\end{Definition}
A detailed discussion of these functionals, and the issue of imposing boundary conditions on $(A,E)$, can be found in \cite{ThiemannQuantumSpinDynamics7}, where a regularised Poisson (even symplectic) structure on the truncated phase spaces $\Gamma_{l}=T^{*}G^{\times |E(\gamma)|},\ l=(\gamma,P_{\gamma},\Pi_{\gamma})\in\cL,$ compatible with \eqref{eq:poisson} is derived, as well:
\begin{Proposition}[cf. \cite{ThiemannQuantumSpinDynamics7}, Theorem 3.2]
\label{prop:loopphasespacepoisson}
Let $l=(\gamma,P_{\gamma},\Pi_{\gamma})\in\cL$ and $f,f'\in C^{\infty}(G)$, then the regularised Poisson structure\index{structure!Poisson!for truncated \acrshort{lqg}-type models|textbf} on $\Gamma_{l}$ w.r.t. to the functionals \eqref{eq:loopphasespacefunctionals} agrees with the Poisson structure on (smooth) polynomial symbols coming from the canonical symplectic form on $T^{*}G^{\times|E(\gamma)|}\cong(G\times\fg)^{\times|E(\gamma)|}$ (cp. theorem III.14 and equations (3.61), (3.62) \& (3.65) of our second article\cite{StottmeisterCoherentStatesQuantumII}) on , i.e.:
\begin{align}
\label{eq:loopphasespacepoisson}
\{f(g_{e}(\ .\ ;\sigma)),f'(g_{e'}(\ .\ ,\sigma))\}_{\Gamma_{l}}(A,E) & = 0, \\[0.25cm] \nonumber
\{P^{e}_{X}(\ .\ ,\ .\ ;\sigma),f'(g_{e'}(\ .\ ,\sigma))\}_{\Gamma_{l}}(A,E) & = \delta^{e,e'}(R_{X}f')(g_{e'}(A;\sigma)), \\[0.25cm] \nonumber
\{P^{e}_{X}(\ .\ ,\ .\ ;\sigma),P^{e'}_{Y}(\ .\ ,\ .\ ;\sigma)\}_{\Gamma_{l}}(A,E) & = -\delta_{e,e'}P^{e}_{[X,Y]}(A,E;\sigma).
\end{align}
\end{Proposition}
Since the functionals \eqref{eq:loopphasespacefunctionals} provide coordinates for the truncated phase spaces $\Gamma_{l},\ l\in\cL,$ we may extend the Poisson structure \eqref{eq:loopphasespacepoisson} to $C^{\infty}(T^{*}G^{|E(\gamma)|})\cong C^{\infty}(T^{*}G)^{\hat{\otimes}|E(\gamma)|},\ \gamma\in l$ (the isomorphism is to be understood in the nuclear Fr\'echet space topology, cf. \cite{AmannVectorValuedDistributions}).\\
The functionals \eqref{eq:loopphasespacefunctionals} behave naturally w.r.t. to composition and inversion of edges\index{edge!composition}\index{edge!inversion}:
\begin{Lemma}
\label{lem:loopphasespacegroupoidrelations}
Let $e=e_{2}\circ e_{1}$, i.e. $e(t)=e_{1}(2t)$, $t\in[0,\tfrac{1}{2}],\ e(t)=e_{2}(2t-1)$, $t\in[\tfrac{1}{2},1]$, and $e^{-1}(t) = e(1-t)$, $t\in[0,1]$, then we have:
\begin{align}
\label{eq:loopphasespacegroupoidrelations}
g_{e}(A;\sigma)(s,t) & = \left\{\begin{matrix*}[l] g_{e_{1}}(A;\sigma)(2s,2t) & : s\leq t\leq\tfrac{1}{2} \\ g_{e_{2}}(A;\sigma)(0,2t-1)g_{e_{1}}(A;\sigma)(2s,1) & : s\leq\tfrac{1}{2}\leq t \\ g_{e_{2}}(A;\sigma)(2s-1,2t-1) & : \tfrac{1}{2}\leq s\leq t \end{matrix*}\right., \\[0.25cm] \nonumber
g_{e^{-1}}(A;\sigma)(s,t) & = g_{e}(A;\sigma)(1-t,1-s)^{-1}
\end{align}
and
\begin{align}
\label{eq:loopphasespacemomentuminversion}
P^{e^{-1}}_{X}(A,E;\sigma) & = -P^{e}_{\textup{Ad}_{g_{e}(A;\sigma)}(X)}(A,E;\sigma).
\end{align}
\begin{Proof}
\eqref{eq:loopphasespacegroupoidrelations} is a simple consequence of the properties of the holonomy map of a connection $A\in\mathcal{A}_{\uP}$. \eqref{eq:loopphasespacemomentuminversion} follows from \eqref{eq:loopphasespacegroupoidrelations}, if we assume that the polyhedronal decomposition of an edge-inverted graph $\gamma^{-1}$ (corresponding to $\gamma$) is chosen s.t. $S_{e^{-1}}=S_{e}$ carries the orientation opposite to $S_{e}$, and the systems of paths satisfy $\rho_{e^{-1}}(x) = \rho_{e}(x),\ x\in S_{e},\ s_{e^{-1}}=1-s_{e}$.
\end{Proof}
\end{Lemma}
The behaviour under the group of gauge transformations $\cG_{\uP}$ is natural as well and even vertex local\cite{StottmeisterStructuralAspectsOf}.\index{Ashtekar-Barbero!variables!behaviour under gauge transformations}
\begin{Lemma}
\label{lem:loopphasespacegauge}
Let $\lambda\in\cG_{\uP}$, and denote by $f_{\lambda}\in C(\uP,G)_{\alpha}$ the corresponding $\alpha$-equivariant $G$-valued function on $\uP$ ($\alpha$ is the conjugation in $G$). Then, the transformations,
\begin{align}
\label{eq:loopphasespacegaugecalssical}
A & \mapsto(\lambda^{-1})^{*}A, \\[0.25cm] \nonumber
E & \mapsto\lambda_{*}E,
\end{align}
where $E$ is identified with its $\textup{Ad}^{*}$-equivariant extension to $\uP$, induce the following transformations on the functionals \eqref{eq:loopphasespacefunctionals}:
\begin{align}
\label{eq:loopphasespacegauge}
g_{e}((\lambda^{-1})^{*}A;\sigma)(s,t) & = f_{\lambda}(\sigma(e(t)))g_{e}(A;\sigma)f_{\lambda}(\sigma(e(s)))^{-1}, \\[0.25cm] \nonumber
P^{e}_{X}((\lambda^{-1})^{*}A,\lambda_{*}E;\sigma) & = P^{e}_{\textup{Ad}_{f_{\lambda}(\sigma(e(1)))^{-1}}(X)}(A,E;\sigma).
\end{align}
\begin{Proof}
This is a simple application of the transformation behaviour of $(A,E)\in|\Lambda|^{1}T^{*}\mathcal{A}_{\uP}$.
\end{Proof}
\end{Lemma}
Let us add an extend remark concerning the structure of the functionals \eqref{eq:loopphasespacefunctionals}, and their dependence on the choice of an auxiliary (measurable) section $\sigma:\Sigma\rightarrow\uP$.
\begin{Remark}
\label{rem:loopphasespacefunctionals}
The space of connections $\mathcal{A}_{\uP}$ can be modelled as an affine space on the space of $\textup{Ad}$-equivariant, horizontal, $\fg$-valued 1-forms on $\uP$, $\overline{\Lambda}^{1}(\uP,\fg)_{\textup{Ad}}$, which serves as configuration space in the Ashtekar-Barbero formulation of general relativity (and also in higher dimensional generalisations, cf. \cite{BodendorferNewVariablesFor1, BodendorferNewVariablesFor2}). The momentum variables, on the other hand, are elements of  $\Gamma\left(T\Sigma\otimes\textup{Ad}^{*}(\uP)\otimes|\Lambda|^{1}(\Sigma)\right)$. But, to make the (densitiesed) cotangent bundle structure explicit, i.e. $(A,E)\in|\Lambda|^{1}T^{*}\mathcal{A}_{\uP}$, we have to identify $\Gamma\left(T\Sigma\otimes\textup{Ad}^{*}(\uP)\right)$ with $T^{*}_{A}\mathcal{A}_{\uP}=(\overline{\Lambda}^{1}(\uP,\fg)_{\textup{Ad}})^{*}$, which, indeed, is possible because $(\overline{\Lambda}^{1}(\uP,\fg)_{\textup{Ad}})^{*}$ is isomorphic with the space of $\textup{Ad}^{*}$-equivariant, horizontal (w.r.t. $A$), $\fg$-valued vector fields on $\uP$, $\overline{\mathfrak{X}}(\uP,\fg)_{\textup{Ad}^{*}}$. Thus, $E$ can be realised as an element of $|\Lambda|^{1}T^{*}_{A}\mathcal{A}_{\uP}$, and does depend on the base point $A$.\\[0.1cm]
Therefore, the impression that the use of the position functionals $g_{e}(A;\sigma)$ is the sole source of dependence of the momentum functional $P^{e}_{X}(A,E;\sigma)$ on $A$ is only apparent, because the usual definition of $(A,E)$ employs the trivialisation $|\Lambda|^{1}T^{*}\mathcal{A}_{\uP}\cong\mathcal{A}_{\uP}\times\Gamma\left(T\Sigma\otimes\textup{Ad}^{*}(\uP)\otimes|\Lambda|^{1}(\Sigma)\right)$, which is structurally similar to the trivialisation $T^{*}G\cong G\times\fg^{*}$, shrouding the (densitiesed) cotangent bundle structure.\\
This said, we may appreciate the special form of the Poisson algebra \eqref{eq:loopphasespacepoisson} generated by the functionals $\eqref{eq:loopphasespacefunctionals}$, which intertwines the (right) trivialisations of $|\Lambda|^{1}T^{*}\mathcal{A}_{\uP}$ and $T^{*}G^{\times|E(\gamma)|}$.\\[0.1cm]
Now, the dependence of the functionals \eqref{eq:loopphasespacefunctionals} on the (measurable) section\index{section!reference} $\sigma:\Sigma\rightarrow\uP$ remains to be clarified:\\
Clearly, if we are given two sections $\sigma, \sigma':\Sigma\rightarrow\uP$, the fibre-transitive (right) action of $G$ on $\uP$ will provide us with a measurable function $g:\Sigma\rightarrow G$, s.t. $\sigma'(x)=R_{g(x)}(\sigma(x))$. Inspecting the definitions \eqref{eq:loopphasespacefunctionals} closely, and making use of the equivariance of the constructions, we find:
\begin{align}
\label{eq:generalisedgauge}
g_{e}(A;\sigma')(s,t) & = g(e(t))^{-1}g_{e}(A;\sigma)(s,t)g(e(s)), \\[0.25cm] \nonumber
P^{e}_{X}(A,E;\sigma') & = P^{e}_{\textup{Ad}_{g(e(1))}(X)}(A,E;\sigma).
\end{align}
Thus, comparing \eqref{eq:loopphasespacegauge} and \eqref{eq:generalisedgauge}, we see that a change of section from $\sigma$ to $\sigma'$ is similar in effect to a gauge transformation, which could also be inferred from the observation that $\lambda\circ\sigma$ defines a section of $\uP$ for $\lambda\in\cG_{\uP}$. But, although these operations of changing the section $\sigma$ and acting with gauge transformation $\lambda$ effect the functionals \eqref{eq:loopphasespacefunctionals} in a similar fashion, they are strictly speaking not equivalent, because the transformations induced by changes of sections are only measurable, while the gauge transformations come with additional regularity properties (semi-analytic in our case), which are influenced by the possible non-triviality of the bundle $\uP$. Nevertheless, as long as we are concerned with a finite collection of structured graphs, or at least locally finite collections\footnote{Unfortunately, this excludes fractal graphs, which are sometimes assumed to be of relevance for loop quantum gravity.}, and the regularity properties allow for a suitable localisation of gauge transformations, e.g. semi-analyticity, the overall effect of the gauge group $\cG_{\uP}$ accounts for all possible changes of sections $\sigma$, as well, due to the vertex local character of \eqref{eq:generalisedgauge} and \eqref{eq:loopphasespacegauge}.\\
But, if the action of the gauge group $\cG_{\uP}$ is essentially equivalent to the action of all (measurable) maps $g:\Sigma\rightarrow G$, we may wonder, whether the topological (and differential geometric) properties of the bundle $\uP$ are in any way reflected in the quantum theory, and thus if (principal) fibre bundles are important to question in the quantum theory at all. The answer to this question is quite subtle, but it can be shown that non-trivial topological properties of the gauge group $\cG_{\uP}$ (e.g. existence of large gauge transformation) leave an imprint on the structure of the algebra of observables (e.g. $\theta$-sectors) under certain conditions (e.g. chirally coupled fermions with chiral anomaly \cite{MorchioChiralSymmetryBreaking, StottmeisterStructuralAspectsOf}). An observation along similar lines was made by Landsman \cite{LandsmanMathematicalTopicsBetween}, i.e. domains of definition for (unbounded) observables of (quantum) particles coupled to external gauge fields can be affected by topological properties of the (classical) bundle $\uP$.
\end{Remark}
Coming back to Poisson relation \eqref{eq:loopphasespacepoisson}, proposition \ref{prop:loopphasespacepoisson} tells us that it makes sense to identify the functional $P^{e}_{(\ .\ )}(\ .\ ,\ .\ ;\sigma)$ with the momentum map of the strongly Hamiltonian left $G$-action $L^{*}_{(\ .\ )^{-1}}$ on the $e$-th component of $\Gamma_{l}$ (cp. equations (3.61) \& (3.62) of our companion article\cite{StottmeisterCoherentStatesQuantumII}). Furthermore, the behaviour under edge inversion \eqref{eq:loopphasespacemomentuminversion}, $e\mapsto e^{-1}$ ($l\mapsto l'$ in $\cL$), is precisely such that it turns $P^{e^{-1}}_{(\ .\ )}(\ .\ ,\ .\ ;\sigma)$ into the momentum map of the compatible right action $R^{*}_{(\ .\ )^{-1}}$:
\begin{align}
\label{eq:leftrightGactions}
& \{P^{e^{-1}}_{X}(\!\ .\!\ ,\!\ .\!\ ;\sigma),f'(g_{e'^{-1}}(\!\ .\!\ ,\sigma))\}_{\Gamma_{l'}}(A,E) \\ \nonumber
 & = \delta^{e,e'}(R_{X}f')(g_{e'^{-1}}(A;\sigma)) \\ \nonumber
 & = \delta^{e,e'}\frac{d}{dt}_{|t=0}f'(\exp(tX)g_{e'}(A;\sigma)^{-1}) \\ \nonumber
 & = \delta^{e,e'}\frac{d}{dt}_{|t=0}(f'\circ(\!\ .\!\ )^{-1})(g_{e'}(A;\sigma)\exp(-tX)) \\ \nonumber
 & = \delta^{e,e'}\frac{d}{dt}_{|t=0}(f'\circ(\!\ .\!\ )^{-1})(\exp(-t \textup{Ad}_{g_{e'}(A;\sigma)}(X))g_{e'}(A;\sigma)) \\ \nonumber
 & = -\delta^{e,e'}(R_{\textup{Ad}_{g_{e'}(A;\sigma)}(X)}(f'\circ(\ .\ )^{-1}))(g_{e'}(A;\sigma)) \\ \nonumber
 & = \{-P^{e}_{\textup{Ad}_{g_{e}(\ .\ ;\sigma)}(X)}(\!\ .\!\ ,\!\ .\!\ ;\sigma),(f'\circ(\!\ .\!\ )^{-1})(g_{e'}(\!\ .\!\ ,\sigma))\}_{\Gamma_{l}}(A,E).
\end{align}
In the (right) trivialisation $T^{*}G^{|E(\gamma)|}\cong(G\times\fg^{*})^{|E(\gamma)|}$ these actions are given explicitly as:
\begin{align}
\label{eq:truncatedGactions}
L^{*}_{h^{-1}}(\theta,g) & = (\textup{Ad}^{*}_{h}(\theta),hg), & R^{*}_{h^{-1}}(\theta,g) & = (\theta,gh^{-1}),
\end{align}
where $g,h\in G,\ \theta\in\fg^{*}$. This, in turn, allows us to associate with each edge $e$ of $\gamma\in l$ the $C^{*}$-dynamical system $(C(G), G, \alpha_{\uL})$, which is determined by the integrated form of \eqref{eq:loopphasespacepoisson} (see equations (3.8) of our second article\cite{StottmeisterCoherentStatesQuantumII}), and thus the transformation group $C^{*}$-algebra $C(G)\rtimes_{\uL}G$. The edge inversion fits into this (global) picture in the following sense:
\begin{Proposition}
\label{prop:loopphasespaceedgeinversion}
Given a structured graph $l\in\cL$, we consider the collection of $C^{*}$-dynamical systems $(C(G), G, \alpha_{\uL})_{e}$, $e\in E(\gamma),$ associated with $\Gamma_{l}$ via the functionals \eqref{eq:loopphasespacefunctionals}. The edge inversion, $e\mapsto e^{-1}$, induces the isomorphism
\begin{align}
\label{eq:edgeinversioniso}
(C(G), G, \alpha_{\uL})_{e^{-1}} & \cong (C(G), G, \alpha_{\uR})_{e},\ \ \ e\in E(\gamma).
\end{align}
Additionally, we have a natural isomorphism $(C(G), G, \alpha_{\uL})\cong(C(G), G, \alpha_{\uR})$, which induces an isomorphism of the assignment $e\mapsto(C(G), G, \alpha_{\uL})_{e}$ form the edge inversion.\\
From the collection of $C^{*}$-dynamical systems $(C(G),G,\alpha_{\uL})_{e},\ e\in E(\gamma),$ we may form the tensor product of the associated transformation group $C^{*}$-algebras, $\mathfrak{A}_{l}:=(C(G)\rtimes_{\uL}G)^{\otimes|E(\gamma)|}$ (which is unambiguous, because $C(G)\rtimes_{\uL}G$ is nuclear (cf. \cite{WilliamsCrossedProductsOf}, $G$ is amenable), and the order of the tensor factors is irrelevant, because associativity and commutativity for the tensor product, e.g. the spatial tensor product, are implemented by natural isomorphisms (cf. \cite{RaeburnMoritaEquivalenceAnd})). The latter satisfies:
\begin{align}
\label{eq:loopphasespacecstar}
(C(G)\rtimes_{\uL}G)^{\otimes|E(\gamma)|} & \cong C(C_{l})\rtimes_{\uL}C_{l}.
\end{align}
\begin{Proof}
The isomorphism \eqref{eq:edgeinversioniso} is immediate from the behaviour of the functionals \eqref{eq:loopphasespacefunctionals} under edge inversion and the comment preceding the proposition. Thus, we only need to prove the isomorphism $(C(G), G, \alpha_{\uL})\cong(C(G), G, \alpha_{\uR})$. We do this via the natural left and right regular integrated representations on $L^{2}(G)$,
\begin{align}
\label{eq:rightleftregularint}
\left(\rho_{\uL}(f)\Psi\right)(g) & = \int_{G}dh\ f(h,g)\Psi(h^{-1}g), & \left(\rho_{\uR}(f)\Psi\right)(g) & = \int_{G}dh\ f(h,g)\Psi(gh)
\end{align}
for $f\in C(G,C(G)), \Psi\in L^{2}(G)$, which provide isomorphisms of the transformation group $C^{*}$-algebras $C(G)\rtimes_{\uL}G$ and $C(G)\rtimes_{\uR}G$ with $\mathcal{K}(L^{2}(G))$ (see definition II \& theorem II.7 of our companion article\cite{StottmeisterCoherentStatesQuantumII}).\\[0.1cm]
The algebras $C^{*}$-algebras $C(G)\rtimes_{\uL}G$ and $C(G)\rtimes_{\uR}G$ are defined as completions of $C(G,C(G))$ in the universal $C^{*}$-norm together with an involution and a convolution product (see definition II.4 of our companion article\cite{StottmeisterCoherentStatesQuantumII}), which involve the left respectively right action of $G$ on itself. We prove the isomorphism by providing an isomorphism of $C(G,C(G))$ that intertwines these structures via $\rho_{\uL}$ and $\rho_{\uR}$.
\begin{align}
\label{eq:l1inversioniso}
\forall f\in C(G,C(G)):\ \ \ I(f)(h,g) & := f(\alpha_{g}(h^{-1}),g)=f(gh^{-1}g^{-1},g), \\ \nonumber
 I^{-1}(f)(h,g) &\ =f(\alpha_{g^{-1}}(h^{-1}),g),\ \ \ h,g\in G.
\end{align}
Clearly, $I:C(G,C(G))\rightarrow C(G,C(G))$ is an isomorphism (with inverse $I^{-1}$), because group multiplication and inversion are continuous. Next, let us see how $\rho_{\uL}$ and $\rho_{\uR}$ are related via $I$:
\begin{align}
\label{eq:rightleftregulariso}
\left(\rho_{\uR}(I(f))\Psi\right)(g) & = \int_{G}dh\ I(f)(h,g)\Psi(gh) = \int_{G}dh\ f(\alpha_{g}(h^{-1}),g)\Psi(gh) \\ \nonumber
 & = \int_{G}dh\ f(h^{-1},g)\Psi(hg) = \int_{G}dh\ f(h,g)\Psi(h^{-1}g) \\ \nonumber
 & = \left(\rho_{\uL}(f)\Psi\right)(g).
\end{align}
The involutions and multiplications are intertwined w.r.t. $I$ as well:
\begin{align}
\label{eq:l1inviso}
\forall f\in C(G,C(G)):\ I\left(f^{*_{\uL}}\right)(h,g) & = f^{*_{\uL}}(\alpha_{g}(h^{-1}),g) \\ \nonumber
 & = \overline{f(\alpha_{g}(h),\alpha_{g}(h)g)} \\ \nonumber
 & = \overline{f(\alpha_{gh}(h),gh)} \\ \nonumber
 & = \overline{f((\alpha_{gh}(h^{-1}))^{-1},gh)} \\ \nonumber
 & = \overline{I(f)(h^{-1},gh)} \\ \nonumber
 & = I(f)^{*_{\uR}}(h,g),
\end{align}
\begin{align}
\label{eq:l1multiso}
 \forall f.f'\in C(G,C(G)):\ I\left(f\ast_{\uL}f'\right)(h,g) & = (f\ast_{\uL}f')(\alpha_{g}(h^{-1}),g) \\ \nonumber
 & = \int_{G}dk\ f(k,g)f'(k^{-1}\alpha_{g}(h^{-1}),k^{-1}g) \\ \nonumber
 & = \int_{G}dk\ f(\alpha_{g}(k),g)f'(\alpha_{g}(k^{-1}h^{-1}),gk^{-1}) \\ \nonumber
 & = \int_{G}dk\ f(\alpha_{g}(k^{-1}),g)f'(\alpha_{gk}((k^{-1}h)^{-1}),gk) \\ \nonumber
 & = \int_{G}dk\ I(f)(k,g)I(f')(k^{-1}h,gk) \\ \nonumber
 & = \left(I(f)\ast_{\uR}I(f')\right)(h,g).
\end{align}
\eqref{eq:loopphasespacecstar} follows from the isomorphisms $C(G)\rtimes_{\uL}G\cong\mathcal{K}(L^{2}(G))$ and $\mathcal{K}(L^{2}(G))^{\otimes 2}\cong\mathcal{K}(L^{2}(G^{\times 2}))$ (cf. \cite{RaeburnMoritaEquivalenceAnd}).
\end{Proof}
\end{Proposition}
\begin{Remark}
\label{rem:globalcotangentstructure}
Noteworthy, the isomorphism \eqref{eq:edgeinversioniso} (or its inverse) reflects the cotangent bundle structure $T^{*}G\cong G\times\fg^{*}$ on the (global) level of $G\times G$, because it is related to the momentum maps  (in the right trivialisation, see equations (3.61) \& (3.44) of our second article\cite{StottmeisterCoherentStatesQuantumII}):
\begin{align}
\label{eq:globalcotangentstructure}
f(h,g) & = f(\exp(X_{h}),g), \\ \nonumber
  & = \int_{\fg^{*}}\!\frac{d\theta}{(2\pi)^{n}}e^{i\theta(X_{h})}\hat{f}^{1}_{\exp}(\theta,g), \\ \nonumber
 & = \int_{\fg^{*}}\!\frac{d\theta}{(2\pi)^{n}}e^{i\theta(X_{h})}\hat{f}^{1}_{\exp}\!\left(\!J^{L^{*}_{(\ .\ )^{-1}}}(\theta,g),g\!\right), \\[0.25cm] \nonumber
 I(f)(h,g^{-1}) & = f(\alpha_{g^{-1}}(h^{-1}),g^{-1}) \\ \nonumber
 & = f(\exp(-\textup{Ad}_{g^{-1}}(X_{h})),g^{-1}) \\ \nonumber
 & = \int_{\fg^{*}}\!\frac{d\theta}{(2\pi)^{n}}e^{i\theta(X_{h})}\hat{f}^{1}_{\exp}(-\textup{Ad}^{*}_{g^{-1}}(\theta),g^{-1}) \\ \nonumber
 & = \int_{\fg^{*}}\!\frac{d\theta}{(2\pi)^{n}}e^{i\theta(X_{h})}\hat{f}^{1}_{\exp}\!\left(\!J^{R^{*}_{(\ .\ )^{-1}}}(\theta,g),g^{-1}\!\right),
\end{align}
where $f_{\exp}(X,g)=f(\exp(X),g)$ and $\exp(X_{h})=h$.
\end{Remark}
So far, we have only analysed the relations between the functionals \eqref{eq:loopphasespacefunctionals} associated with structured graphs $l,l'\in\cL$ that are related via edge inversion, but it is possible to introduce partial orders, $\leq$ and $\lesssim$, on $\cL$ that leads to a projective structure on the collection of truncated phase spaces $\Gamma_{l},\ l\in\cL$ (see below). $\!\leq$ turns out to be compatible with the Poisson algebra \eqref{eq:loopphasespacepoisson} and certain generalisations of the $C^{*}$-dynamical systems introduced in proposition \ref{prop:loopphasespaceedgeinversion}. This will also explain, why we have not made the dependence of the functionals \eqref{eq:loopphasespacefunctionals} on $l\in\cL$ explicit, but only indicated a dependence on $\gamma\in l$.
\begin{Definition}[cp. \cite{ThiemannQuantumSpinDynamics7} \& \cite{LaneryProjectiveLoopQuantum}]
\label{def:loopphasespacepartialorder}\index{graph!equivalence|textbf}
Given two structured graphs $l=(\gamma,P_{\gamma},\Pi_{\gamma}),l'=(\gamma',P'_{\gamma'},\Pi'_{\gamma'})\in\cL$, we say that $l\leq l'$ if $\gamma\subseteq\gamma'$, i.e. the oriented graph $\gamma\in l$ is an oriented subgraph of the oriented graph $\gamma'\in l'$.\\
If $l\leq l'$ and $l'\leq l$, we say that $l$ and $l'$ are equivalent, $l\sim l'$.\\
Alternatively, we say that $l\lesssim l'$, if $|\gamma|\subseteq|\gamma'|$ (the non-oriented graphs underlying $\gamma$ and $\gamma'$ agree). If $l\lesssim l'$ and $l'\lesssim l$, we say that $l$ and $l'$ are equivalent up to orientation, $l\simeq l'$.\\[0.1cm]
Two other, but somewhat different, partial orders are the following (cf. \cite{LaneryProjectiveLoopQuantum}, p. 52-53):\\[0.1cm]
We say $l\lessdot_{\uL} l'$, if $|\gamma|\subseteq|\gamma'|$, and
\begin{align}
\label{eq:graphendsegequivleft}
\forall e\in E(\gamma): \exists e'\in E(\gamma'): \exists s_{0}\in[0,1): e_{|[s_{0},1]} & = e'.
\end{align}
We write, $l\doteq_{\uL} l'$, if $l\lessdot_{\uL} l'$ and $l'\lessdot _{\uL}l$.\\
We say $l\lessdot_{\uR} l'$, if $|\gamma|\subseteq|\gamma'|$, and
\begin{align}
\label{eq:graphendsegequivright}
\forall e\in E(\gamma): \exists e'\in E(\gamma'): \exists s_{0}\in(0,1]: e_{|[0,s_{0}]} & = e'.
\end{align}
We write, $l\doteq_{\uR} l'$, if $l\lessdot_{\uR} l'$ and $l'\lessdot _{\uR}l$.\\
Clearly, $l\doteq_{\uL} l'$ or $l\doteq_{\uR} l'$, if and only if $\gamma=\gamma'$ ($s_{0}=0$ is necessary).
Loosely speaking, $\lessdot_{\uL}$ and $\lessdot_{\uR}$ encode the condition that a graph $\gamma'$, which is finer (and possibly larger) than another graph $\gamma$, contains an (oriented) edge $e'\in E(\gamma')$ corresponding to the last respectively first part of an (oriented) edge $e\in E(\gamma)$.
\end{Definition}
It follows from the discussion in \cite{ThiemannQuantumSpinDynamics7} that $(\cL,\leq)$ and $(\cL,\lesssim)$ are partially ordered sets\footnote{The partial order $\leq$ is essentially the one, $\prec$, defined in \cite{ThiemannQuantumSpinDynamics7}.}. Moreover, $(\cL,\lesssim)$ is directed, in contrast with $(\cL,\leq)$, which follows, because any two non-oriented, finite, semi-analytic graphs $\gamma,\gamma'$ have a common refined graph $\gamma''$, that admits an orientation and a dual polyhedronal decomposition (cf. \cite{ThiemannQuantumSpinDynamics7}, see also \cite{ThiemannModernCanonicalQuantum}, Section 6.2.2).\\
It is also easy to see that $(\cL,\lessdot_{\uL})$ and $(\cL,\lessdot_{\uR})$ are partially ordered and directed (cf. \cite{LaneryProjectiveLoopQuantum}), because $\gamma''$ can be oriented s.t. \eqref{eq:graphendsegequivleft} respectively \eqref{eq:graphendsegequivright} are satisfied w.r.t. the edge sets $E(\gamma)$, $E(\gamma'')$ and $E(\gamma')$, $E(\gamma'')$.\\
In case we are not in danger of ambiguities, we will use $\cL$ as a short hand for $(\cL,\leq)$, $(\cL,\lesssim)$, $(\cL,\lessdot_{\uL})$ and $(\cL,\lessdot_{\uR})$.\\
In the next theorem (\ref{thm:loopphasespacepartialordercompatible}), we show that the partial orders $\leq$, $\lessdot_{\uL}$ and $\lessdot_{\uR}$ on $\cL$ are compatible with the Poisson structures defined on $\Gamma_{l},\ l\in\cL$. We also show, why we have, at this point, to deal, with oriented graphs, $\Gamma^{\textup{sa},\uparrow}_{0}$, although edge inversion $e\rightarrow e^{-1}$ induces and isomorphism of $\Gamma_{l}$ and $\Gamma_{l'}$ (see lemma \ref{lem:loopphasespacegroupoidrelations} \& \eqref{eq:leftrightGactions}), when $\gamma$ and $\gamma'$ agree up to some edge orientations.\\
The reason for this lies in a compatibility condition of edge inversion and composition, that is not necessarily satisfied for the corresponding maps between the truncated phase spaces $\Gamma_{l},\ l\in\cL$\footnote{It seems that this non-trivial condition has been overlooked in the main part of the literature with the exception of \cite{LaneryProjectiveLoopQuantum}, where the partial orders $\lessdot_{\uL}$ and $\lessdot_{\uR}$ are defined making the use of edge inversions obsolete at the phase space level..}.\\
In contrast, the Ashtekar-Isham-Lewandowski Hilbert space, $L^{2}(\overline{\mathcal{A}})$, which is a fundamental building block of loop quantum gravity, arises from a projective structure constructed w.r.t. finite, non-oriented, semi-analytic graphs $\Gamma^{\textup{sa}}_{0}$ instead of $\cL$ (cf. \cite{ThiemannModernCanonicalQuantum} for a general exposition, and original references). Thus, there seems to be a certain tension between the phase space quantisation for loop quantum gravity presented here (cf. \cite{ThiemannQuantumSpinDynamics7, ThiemannGaugeFieldTheory1, ThiemannGaugeFieldTheory2, ThiemannGaugeFieldTheory2, ThiemannGaugeFieldTheory4, FreidelContinuousFormulationOf}), and the framework based on the holonomy-flux algebra and its Hilbert space representation on $L^{2}(\overline{\mathcal{A}})$.\\[0.1cm]
The link between the two can be roughly understood as follows:\\
If we consider only the images of the holonomy functionals \eqref{eq:loopphasespacefunctionals}, we will obtain the truncated configuration spaces $C_{l}\cong G^{|E(\gamma)|},\ l\in\cL,$ which admit the coarsening $\lesssim$ of the partial order $\leq$ introduced above. Then, we will have projections $p_{ll'}:C_{l'}\rightarrow C_{l}$, if $l\lesssim l'$, which are compatible with $\lesssim$, i.e. $p_{ll''}=p_{ll'}\circ p_{l'l''}$ for $l\lesssim l'\lesssim l''$. Furthermore, these maps can be lifted to symplectic projections $\tilde{p}_{ll'}:\Gamma_{l'}\rightarrow\Gamma_{l}$, but these lifts are not unique, and therefore turn out to be only compatible with $\leq$, $\lessdot_{\uL}$ and $\lessdot_{\uR}$ instead of $\lesssim$. \\
We will further comment on the implications of this issue on the relation between phase space quantisation and holonomy-flux algebras in theorem \ref{thm:loopphasespacequant} and the outlook \ref{sec:con}.
\begin{Theorem}
\label{thm:loopphasespacepartialordercompatible}
Given $l,l'\in\cL$, s.t. $l\lesssim l'$, we have smooth projections $\tilde{p}^{c}_{ll'}:\Gamma_{l'}\rightarrow\Gamma_{l}$, $c=\{c_{e'}(l,l')\}_{e'\in E(\gamma')}\subset\R$, defined by
\begin{align}
\label{eq:loopphasespaceprojection}
g_{e}(A;\sigma) & = g_{e'_{m}}(A;\sigma)^{s_{m}}...g_{e'_{1}}(A;\sigma)^{s_{1}}, \\ \nonumber
P^{e}_{X}(A,E;\sigma) & = c_{e'_{m}}(l,l')P^{e'^{s_{m}}_{m}}_{X}(A,E;\sigma)+...+c_{e'_{1}}(l,l')P^{e'^{s_{1}}_{1}}_{\textup{Ad}_{(g_{e'_{m}}(A;\sigma)^{s_{m}}...g_{e'_{2}}(A;\sigma)^{s_{2}})^{-1}}(X)}(A,E;\sigma),\\ \nonumber \sum_{n=1}^{m}c_{e'_{n}}(l,l') & =1,
\end{align}
where $e\in E(\gamma),\ e'_{1},...,e'_{m}\in E(\gamma')$ s.t. $e=e'^{s_{m}}_{m}\circ ... \circ e'^{s_{1}}_{1}$ with $s_{n}\in\{\pm1\}\ \forall\ n=1,...,m$, because $|\gamma|\subseteq|\gamma'|$.\\[0.1cm]
If $c_{e'_{n}}=1$ for $n=1$ or $n=m$ in the decomposition of an edge $e=e'^{s_{m}}_{m}\circ...\circ e'^{s_{1}}_{1}$, we denote the corresponding maps by $p^{\uR}_{ll'}$ and $p^{\uL}_{ll'}$, which have the properties:
\begin{itemize}
\item[1.] The dual maps $\tilde{p}^{\uR,\uL*}_{ll'}:C^{\infty}(\Gamma_{l})\rightarrow C^{\infty}(\Gamma_{l'})$ are injective, continuous Poisson maps w.r.t. \eqref{eq:loopphasespacepoisson}:
\begin{align}
\label{eq:loopphasespaceinjection}
\forall f,f'\in C^{\infty}(\Gamma_{l}):\ \tilde{p}^{\uR,\uL *}_{ll'}\{f,f'\}_{\Gamma_{l}} & = \{\tilde{p}^{\uR,\uL *}_{ll'}f,\tilde{p}^{\uR,\uL *}_{ll'}f'\}_{\Gamma_{l'}}.
\end{align}
\item[2.] In case, $l$ is equivalent to $l'$ ($l\simeq l'$), $\tilde{p}^{\uR *}_{ll'}=\tilde{p}^{\uL *}_{ll'}$ and $\tilde{p}^{\uR *}_{l'l}=\tilde{p}^{\uL *}_{l'l}$, are Poisson isomorphisms that are inverse to one another.
\item[3.] If $l\leq l'\leq l''$, the maps $\tilde{p}^{\uR,\uL}_{ll'},\ \tilde{p}^{\uR,\uL}_{l'l''}$ and $\tilde{p}^{\uR,\uL}_{ll''}$ are compatible with transitivity of $\leq$, i.e. $\tilde{p}^{\uR,\uL}_{ll'}\circ \tilde{p}^{\uR,\uL}_{l'l''}=\tilde{p}^{\uR,\uL}_{ll''}$ (edge orientations coincide, $\forall n=1,...,m : s_{n}=1$).\\
If $l\lessdot_{\uR,\uL} l'\lessdot_{\uR,\uL} l''$, the maps $\tilde{p}^{\uR,\uL}_{ll'},\ \tilde{p}^{\uR,\uL}_{l'l''}$ and $\tilde{p}^{\uR,\uL}_{ll''}$ are compatible with transitivity of $\leq$, i.e. $\tilde{p}^{\uR,\uL}_{ll'}\circ \tilde{p}^{\uR,\uL}_{l'l''}=\tilde{p}^{\uR,\uL}_{ll''}$ (edge orientations coincide for $n=m$ respectively $n=1$, i.e. $s_{m}=1$ or $s_{1}=1$).
\item[4.] If $\gamma\subset\gamma'$ and $l^{-1}, l'^{-1}$ denote the structured graphs with all edge orientations reversed, and \\ $p_{l^{-1}l}:\Gamma_{l}\rightarrow\Gamma_{l^{-1}},\ l\in\cL,$ are the edge inversion maps ($\tilde{p}^{\uR}_{l^{-1}l}=\tilde{p}^{\uL}_{l^{-1}l}$), we have:
\begin{align}
\label{eq:edgeinversioncompatibilitydiagram}
\xymatrix{
\Gamma_{l'} \ar[rr]^{\tilde{p}_{l'^{-1}l'}} \ar[d]_{\tilde{p}^{\uL}_{ll'}} &  & \Gamma_{l'^{-1}} \ar[d]^{\tilde{p}^{\uR}_{l^{-1}l'^{-1}}} \\
\Gamma_{l} \ar[rr]_{\tilde{p}_{l^{-1}l}} &  & \Gamma_{l^{-1}}
}
\end{align}
\end{itemize}
In general, $p^{c *}_{ll'},\ l\leq l'$ will only be Poisson for those choices of $c$, s.t. for every composition $e=e'^{s_{m}}_{m}\circ ... \circ e'^{s_{1}}_{1}$, we have $c_{e'_{n}}(l,l')=1$ for some $n=1,...,m$ (all other $c_{e'_{n}}(l,l')$'s vanish).
\begin{Proof}
Let us first explain, why the maps $\tilde{p}^{c}_{ll'}$ are natural lifts of the maps $p_{ll'}$ (the latter arise from the holonomy part of \eqref{eq:loopphasespaceprojection}), i.e.
\begin{align}
\label{eq:loopphasespacelift}
\xymatrix{
\Gamma_{l'} \ar[r]^{\tilde{p}^{c}_{ll'}} \ar[d]_{\varpi_{l'}} & \Gamma_{l} \ar[d]^{\varpi_{l}} \\
C_{l'} \ar[r]_{p_{ll'}}  & C_{l}
}
\end{align}
If we consider a function $f$ on $G_{e}$, where $e\in E(\gamma)$ decomposes in $\gamma'$ as $e=e'^{s_{m}}_{m}\circ ... \circ e'^{s_{1}}_{1}$ for some $e'_{1},..,e'_{m}\in E(\gamma')$, we can pull it back to $\bigtimes^{m}_{n=1}G_{e'_{n}}$ via $p_{ll'}$ (to this end, we extend $f$ by $1$ on the other copies of $G$ in $\Gamma_{l}$). Especially, we may pull back $R^{e}_{X}f$ for some $X\in\fg_{e}$, where $R^{e}$ is the right invariant derivation on the $e$-th copy of $G$:
\begin{align}
\label{eq:rightivariantpullbackmap}
\left(p^{*}_{ll'}(R_{X}f)\right)(g_{e'_{m}},...,g_{e'_{1}}) & = \frac{d}{dt}_{|t=0}f\Big(e^{tX}g_{e'_{m}}^{s_{m}}...g_{e'_{1}}^{s_{1}}\Big) \\ \nonumber
 & = \frac{d}{dt}_{|t=0}f\Big(e^{tc_{e'_{m}}X}g_{e'_{m}}^{s_{m}}...e^{tc_{e'_{1}}\textup{Ad}_{(g_{e'_{m}}^{s_{m}}...g_{e'_{2}}^{s_{2}})^{-1}}(X)}g_{e'_{1}}^{s_{1}}\Big)\\ \nonumber
 & = \Big(\Big(c_{e'_{m}}R^{e'^{s_{m}}_{m}}_{X}+...+c_{e'_{1}}R^{e'^{s_{1}}_{1}}_{\textup{Ad}_{(g_{e'_{m}}^{s_{m}}...g_{e'_{2}}^{s_{2}})^{-1}}(X)}\Big)(p^{*}_{ll'}f)\Big)(g_{e'_{m}},...,g_{e'_{1}}) 
\end{align}
for $\sum_{n=1}^{m}c_{e'_{m}}=1$. But, since the right invariant derivation $R^{e}$ is generated by the Poisson bracket with $P^{e}$, if we interpret $f$ as a function on $\Gamma_{l}$ via the cotangent bundle projection, we see that $P^{e}$ should arise from the $P^{e'_{n}},\ n=1,...,m,$ in precisely the way given in \eqref{eq:loopphasespaceprojection}.\\[0.1cm]
That $\tilde{p}^{c}_{ll'}:\Gamma_{l'}\rightarrow\Gamma_{l},\ c=\{c_{e'}\}_{e'\in E(\gamma')}$ are projections is obvious from the definition (we suppress the possible dependence of $c$ on $l,l'$ at this point), and the fact, that a subgraph $|\gamma|\subseteq|\gamma'|$ is obtained by removing and composing edges. Smoothness of $\tilde{p}^{c}_{ll'}$ is implied by the smoothness of the group operations.\\[0.1cm]
Next, we show that $\tilde{p}^{c *}_{ll'}$ is Poisson if and only if $c_{e'_{n}}=1$ for some $n=1,...,m$ (all other $c$'s vanish). To deduce this, we realise that any $\tilde{p}^{c}_{ll'},\ l\lesssim l',$ is obtained from the successive application of three fundamental operations:
\begin{itemize}
	\item[1.] Removal of an edge $e'$ from $\gamma'$, i.e. $\tilde{r}:T^{*}G^{\times2}\rightarrow T^{*}G,\ \tilde{r}((\theta_{2},g_{2}),(\theta_{1},g_{1}))=(\theta_{2},g_{2})$.
	\item[2.] Composition of two edges, $e=e_{2}\circ e_{1}$, i.e. $\tilde{p}^{c}:T^{*}G^{\times2}\rightarrow T^{*}G$, \\ $\tilde{p}^{c}((\theta_{2},g_{2}),(\theta_{1},g_{1}))=(c \textup{Ad}^{*}_{g_{2}}(\theta_{1})+(1-c)\theta_{2},g_{2}g_{1})$.
	\item[3.] Inversion of an edge $e\mapsto e^{-1}$, i.e. $\tilde{\iota}:T^{*}G\rightarrow T^{*}G,\ \iota(\theta,g)=(-\textup{Ad}^{*}_{g^{-1}}(\theta),g^{-1})$.
\end{itemize}
It is obvious, that $\tilde{r}^{*}:C^{\infty}(T^{*}G)\rightarrow C^{\infty}(T^{*}G^{\times2})$ is Poisson, i.e.
\begin{align}
\label{eq:removalpoisson}
\forall f,f'\in C^{\infty}(T^{*}G):\ \{\tilde{r}^{*}f,\tilde{r}^{*}f'\}_{T^{*}G^{\times2}}((\theta_{2},g_{2}),(\theta_{1},g_{1})) & = \tilde{r}^{*}\left(\{f,f'\}_{T^{*}G}\right)((\theta_{2},g_{2}),(\theta_{1},g_{1})),
\end{align}
because $\tilde{r}^{*}f$ and $\tilde{r}^{*}f'$ depend only on $(\theta_{2},g_{2})$. A short calculation shows, what conditions on $c$ are implied, if $\tilde{p}^{c *}:C^{\infty}(T^{*}G)\rightarrow C^{\infty}(T^{*}G^{\times2})$ is assumed to be Poisson.
\begin{align}
\label{eq:compositionpoisson}
 & \{\tilde{p}^{c *}\!f,\tilde{p}^{c *}\!f'\}_{T^{*}G^{\times2}}((\theta_{2},g_{2}),\!(\theta_{1},g_{1})) \\ \nonumber
 & = \langle\partial_{\theta_{1}}\tilde{p}^{c *}\!f,R_{1}\tilde{p}^{c *}\!f'\rangle((\theta_{2},g_{2}),\!(\theta_{1},g_{1})) \!-\! \langle\partial_{\theta_{1}}\tilde{p}^{c *}\!f'\!,R_{1}\tilde{p}^{c *}\!f\rangle((\theta_{2},g_{2}),\!(\theta_{1},g_{1})) \\ \nonumber
 &\hspace{0.2cm} - \theta_{1}([\partial_{\theta_{1}}\tilde{p}^{c *}\!f,\partial_{\theta_{1}}\tilde{p}^{c *}\!f']((\theta_{2},g_{2}),(\theta_{1},g_{1}))) \\ \nonumber
 &\hspace{0.2cm} +\langle\partial_{\theta_{2}}\tilde{p}^{c *}\!f,R_{2}\tilde{p}^{c *}\!f'\rangle((\theta_{2},g_{2}),\!(\theta_{1},g_{1})) \!-\! \langle\partial_{\theta_{2}}\tilde{p}^{c *}\!f'\!,R_{2}\tilde{p}^{c *}\!f\rangle((\theta_{2},g_{2}),\!(\theta_{1},g_{1})) \\ \nonumber
 &\hspace{0.2cm} - \theta_{2}([\partial_{\theta_{2}}\tilde{p}^{c *}\!f,\partial_{\theta_{2}}\tilde{p}^{c *}\!f']((\theta_{2},g_{2}),(\theta_{1},g_{1}))) \\ \nonumber
 & = \tilde{p}^{c *}\left(\langle\partial_{\theta}f,Rf'\rangle - \langle\partial_{\theta}f',Rf\rangle\right)((\theta_{2},g_{2}),(\theta_{1},g_{1})) \\ \nonumber
 &\hspace{0.2cm} - (2-c)c \textup{Ad}^{*}_{g_{2}}(\theta_{1})\left((\tilde{p}^{c *}[\partial_{\theta}f,\partial_{\theta}f'])((\theta_{2},g_{2}),(\theta_{1},g_{1}))\right) \\ \nonumber
 &\hspace{0.2cm} - (1-c)^{2}\theta_{2}\left((\tilde{p}^{c *}[\partial_{\theta}f,\partial_{\theta}f'])((\theta_{2},g_{2}),(\theta_{1},g_{1}))\right) \\ \nonumber
 & = \tilde{p}^{c *}\left(\langle\partial_{\theta}f,Rf'\rangle - \langle\partial_{\theta}f',Rf\rangle\right)((\theta_{2},g_{2}),(\theta_{1},g_{1})) \\ \nonumber
 &\hspace{0.2cm} - (2-c)\theta_{\tilde{p}^{c}((\theta_{2},g_{2}),(\theta_{1},g_{1}))}\left((\tilde{p}^{c *}[\partial_{\theta}f,\partial_{\theta}f'])((\theta_{2},g_{2}),(\theta_{1},g_{1}))\right) \\ \nonumber
 &\hspace{0.2cm} + (1-c)\theta_{2}\left((\tilde{p}^{c *}[\partial_{\theta}f,\partial_{\theta}f'])((\theta_{2},g_{2}),(\theta_{1},g_{1}))\right) \\ \nonumber
 & = \tilde{p}^{c *}\left(\{f,f'\}_{T^{*}G}\right)((\theta_{2},g_{2}),(\theta_{1},g_{1})) \\ \nonumber
 &\hspace{0.2cm} - (1-c)\theta_{\tilde{p}^{c}((\theta_{2},g_{2}),(\theta_{1},g_{1}))}\left((\tilde{p}^{c *}[\partial_{\theta}f,\partial_{\theta}f'])((\theta_{2},g_{2}),(\theta_{1},g_{1}))\right) \\ \nonumber
 &\hspace{0.2cm} + (1-c)\theta_{2}\left((\tilde{p}^{c *}[\partial_{\theta}f,\partial_{\theta}f'])((\theta_{2},g_{2}),(\theta_{1},g_{1}))\right) ,
\end{align}
for all $f,f'\in C^{\infty}(T^{*}G^{\times2})$. Here, we used the formula for the canonical Poisson structure on $T^{*}G$ for $\{\ ,\ \}_{T^{*}G}$ and $\{\ ,\ \}_{T^{*}G^{\times2}}$ (see theorem III.14 of our second article\cite{StottmeisterCoherentStatesQuantumII}). The last line shows, that the only possible choices for $c$, to make $\tilde{p}^{c *}$ a Poisson map, are $c=1$ or $c=0$. Clearly, a similar phenomenon occurs for compositions involving more than 2 edges. Even, if we were to relax the condition $\sum_{n=1}^{m}c_{e'_{n}}=1$, this phenomenon would persist. \\
Another short calculation shows that $\tilde{\iota}^{*}:C^{\infty}(T^{*}G)\rightarrow C^{\infty}(T^{*}G)$ is Poisson.
\begin{align}
\label{eq:iotapoisson}
\{\tilde{\iota}^{*}f,\tilde{\iota}^{*}f'\}_{T^{*}G}(\theta,g) & = \langle\partial_{\theta}\tilde{\iota}^{*}f,R\tilde{\iota}^{*}f'\rangle(\theta,g) - \langle\partial_{\theta}\tilde{\iota}^{*}f',R\tilde{\iota}^{*}f\rangle(\theta,g) - \theta([\partial_{\theta}\tilde{\iota}^{*}f,\partial_{\theta}\tilde{\iota}^{*}f'](\theta,g)) \\ \nonumber
 & = \tilde{\iota}^{*}(\langle\partial_{\theta}f,Rf'\rangle)(\theta,g) - \tilde{\iota}^{*}(\langle\partial_{\theta}f',Rf\rangle)(\theta,g) + \textup{Ad}^{*}_{g^{-1}}(\theta)(\tilde{\iota}^{*}[\partial_{\theta}f,\partial_{\theta}f'](\theta,g)) \\ \nonumber
 & = \tilde{\iota}^{*}\{f,f'\}_{T^{*}G}(\theta,g).
\end{align}
If $l$ and $l'$ are equivalent, $l\simeq l'$, $\tilde{p}^{c}_{ll'}=\tilde{p}_{ll'}$ (no $c$-dependence) is induced from the map $\tilde{\iota}$ on single edges. Since $\tilde{\iota}$ is an involution, $\tilde{\iota}\circ\tilde{\iota} = \id_{T^{*}G}$, we conclude that
\begin{align}
\label{eq:equivalentgraphsinversion}
\forall c:\ \forall l\simeq l:\ (\tilde{p}_{ll'})^{-1} & = \tilde{p}_{l'l}.
\end{align}
Moreover, because $\tilde{\iota}^{*}$ is Poisson, $\tilde{p}^{*}_{ll'}$ is Poisson for all $l\simeq l'$.\\[0.1cm]
To understand what conditions are imposed on the set $c=\{c_{e'}\}_{e'\in E(\gamma')}$ by demanding compatibility with transitivity w.r.t. $\leq$, $\lessdot_{\uR}$ or $\lessdot_{\uL}$, we take a look at the implications coming from the associativity of edge composition. This certainly encompasses the case of composing three edges in the forms $(e_{3}\circ e_{2})\circ e_{1}=e_{2'}\circ e_{1}=e$ and $e_{3}\circ(e_{2}\circ e_{1})=e_{3}\circ e_{1'}=e$:
\begin{align}
\label{eq:projectiontransitivity}
& \tilde{p}^{c}_{2'1}:T^{*}G_{2'}\times T^{*}G_{1} \rightarrow T^{*}G, \\ \nonumber
& \tilde{p}^{c}_{2'1}((\theta_{2'},g_{2'}),(\theta_{1},g_{1})) = (c^{2'1}_{1}\textup{Ad}^{*}_{g_{2'}}(\theta_{1})+(1-c^{2'1}_{1})\theta_{2'},g_{2'}g_{1}), \\ \nonumber
& \tilde{p}^{c}_{(32)1}:T^{*}G_{3}\times T^{*}G_{2}\times T^{*}G_{1} \rightarrow T^{*}G_{2'}\times T^{*}G_{1}, \\ \nonumber
& \tilde{p}^{c}_{(32)1}((\theta_{3},g_{3}),(\theta_{2},g_{2}),(\theta_{1},g_{1})) = ((c^{(32)1}_{2}\textup{Ad}^{*}_{g_{2}}(\theta_{2})+(1-c^{(32)1}_{2})\theta_{3},g_{3}g_{2}),(\theta_{1},g_{1})), \\[0.25cm] \nonumber
& \tilde{p}^{c}_{31'}:T^{*}G_{3'}\times T^{*}G_{1} \rightarrow T^{*}G, \\ \nonumber
& \tilde{p}^{c}_{31'}((\theta_{3},g_{3}),(\theta_{2},g_{2}),(\theta_{1},g_{1})) = ((1-c^{31'}_{3})\textup{Ad}^{*}_{g_{3'}}(\theta_{1})+c^{31'}_{3}\theta_{3'},g_{3'}g_{1}), \\ \nonumber
& \tilde{p}^{c}_{3(21)}:T^{*}G_{3}\times T^{*}G_{2}\times T^{*}G_{1} \rightarrow T^{*}G_{3}\times T^{*}G_{1'}, \\ \nonumber
& \tilde{p}^{c}_{3(21)}((\theta_{3},g_{3}),(\theta_{2},g_{2}),(\theta_{1},g_{1})) = ((\theta_{3},g_{3})(c^{3(21)}_{1}\textup{Ad}^{*}_{g_{1}}(\theta_{1})+(1-c^{3(21)}_{1})\theta_{2},g_{2}g_{1})), \\[0.25cm] \nonumber
& \tilde{p}^{c}_{321}:T^{*}G_{3}\times T^{*}G_{2}\times T^{*}G_{1} \rightarrow T^{*}G, \\ \nonumber
& \tilde{p}^{c}((\theta_{3},g_{3}),(\theta_{2},g_{2}),(\theta_{1},g_{1})) = (c^{321}_{1}\textup{Ad}^{*}_{g_{3}g_{2}}(\theta_{1})+c^{321}_{2}\textup{Ad}^{*}_{g_{3}}(\theta_{2})+c^{321}_{3}\theta_{3},g_{3}g_{2}g_{1}),
\end{align}
\begin{align}
\label{eq:projectiontransitivityconditions}
 & \Rightarrow\ \!& (\tilde{p}^{c}_{2'1}\!\circ\! p^{c}_{(32)1})((\theta_{3},g_{3}),\!(\theta_{2},g_{2}),\!(\theta_{1},g_{1}))  & = (\tilde{p}^{c}_{31'}\!\circ\! \tilde{p}^{c}_{3(21)})((\theta_{3},g_{3}),\!(\theta_{2},g_{2}),\!(\theta_{1},g_{1})) \\ \nonumber
 & & & =  \tilde{p}^{c}((\theta_{3},g_{3}),\!(\theta_{2},g_{2}),\!(\theta_{1},g_{1}))\\ \nonumber
 & \Leftrightarrow\ \!& c^{321}_{1} = c^{2'1}_{1} = (1-c^{31'}_{3})c^{3(21)}_{1} & \wedge\ c^{321}_{2} = (1-c^{2'1}_{1})c^{(32)1}_{2} = (1-c^{31'}_{3})(1-c^{3(21)}_{1}) \\ \nonumber
 & & &\wedge\ c^{321}_{3} = (1-c^{2'1}_{1})(1-c^{(32)1}_{2}) = c^{31'}_{3},
\end{align}
where we used the constraints $c^{321}_{1}+c^{321}_{2}+c^{321}_{3}=1,\ c^{2'1}_{1}+c^{2'1}_{2'}=1,\ c^{31'}_{1'}+c^{31'}_{3}=1,\ c^{(32)1}_{2}+c^{(32)1}_{3}=1$ and $c^{3(21)}_{1}+c^{3(21)}_{2}=1$. \\
Since we are only interested in Poisson maps, the only interesting cases to check are $c^{321}_{n}=\delta_{n1}$, $c^{321}_{n}=\delta_{n2}$ and $c^{321}_{n}=\delta_{n3}$, $n=1,2,3$. The first imposes $\tilde{p}^{c}_{2'1}=\tilde{p}^{\uR}_{2'1}$, $\tilde{p}^{c}_{(32)1}=\tilde{p}^{\uR}_{(32)1}$ or $\tilde{p}^{\uL}_{(32)1}$, $\tilde{p}^{c}_{31'}=\tilde{p}^{\uR}_{31'}$, $\tilde{p}^{c}_{3(21)}=\tilde{p}^{\uR}_{3(21)}$. The second case forces us to set $\tilde{p}^{c}_{2'1}=\tilde{p}^{\uL}_{2'1}$, $\tilde{p}^{c}_{(32)1}=\tilde{p}^{\uR}_{(32)1}$, $\tilde{p}^{c}_{31'}=\tilde{p}^{\uL}_{31'}$, $\tilde{p}^{c}_{3(21)}=\tilde{p}^{\uL}_{3(21)}$. The third case gives $\tilde{p}^{c}_{2'1}=\tilde{p}^{\uL}_{2'1}$, $\tilde{p}^{c}_{(32)1}=\tilde{p}^{\uL}_{(32)1}$, $\tilde{p}^{c}_{31'}=\tilde{p}^{\uL}_{31'}$, $\tilde{p}^{c}_{3(21)}=\tilde{p}^{\uL}_{3(21)}$ or $\tilde{p}^{\uR}_{3(21)}$. \\
Thus, we infer that choosing $\tilde{p}^{\uL}_{ll'}$ or $\tilde{p}^{\uR}_{ll'}$ for all $l\leq l'$ generates a system of maps compatible with transitivity of $\leq$, because outer left or right (w.r.t. edge orientation, i.e. $n=1$ or $n=m$ in a composition chain) $\theta$-labels are preserved in composition sequences. This property is not affected by edge removal, because the latter only generates new (left or right) edge boundaries, which must be present in a subgraph independent of the specific sequence of composing and removing edges.\\
An analogous argument works for $\lessdot_{\uR}$ and $\lessdot_{\uL}$ in combination with $\tilde{p}^{\uR}_{ll'}$ and $\tilde{p}^{\uL}_{ll'}$ respectively, because these partial orders preserve the notion of first respectively last part of an edge between an oriented graph and its oriented subgraphs (cf. \cite{LaneryProjectiveLoopQuantum}, p. 52-53).
\\[0.1cm]
The continuity of $\tilde{p}^{\uL *}_{ll'}$ can be reduced to the continuity of $\tilde{p}^{\uL *}\!:\!C^{\infty}(T^{*}G)\rightarrow C^{\infty}(T^{*}G^{\times2})$, \\ $(\tilde{p}^{\uL *}f)((\theta_{1},\theta_{2}),(g_{1},g_{2})):=f(\theta_{2},g_{2}g_{1}),$ which corresponds to the fundamental operation of composing two edges $e=e_{2}\circ e_{1}$.
\begin{align}
\label{eq:loopphasespaceinjectioncontinuity}
||\tilde{p}^{\uL *}f||_{m,(K_{1},K_{2})} & = \sup_{\substack{\alpha_{1},\alpha_{2},\beta_{1},\beta_{2}\in\N^{n}_{0} \\ |\alpha_{1}|+|\alpha_{2}|+|\beta_{1}|+|\beta_{2}|\leq m}}\sup_{\substack{(\theta_{1},\theta2)\in K_{1}\times K_{2} \\ g_{1},g_{2}\in G}}\left|\left(R^{\alpha_{1}}_{1}R^{\alpha_{2}}_{2}\partial^{\beta_{1}}_{\theta_{1}}\partial^{\beta_{2}}_{\theta_{2}}(\tilde{p}^{\uL *}f)\right)((\theta_{1},\theta_{2}),(g_{1},g_{2}))\right| \\ \nonumber
 & = \sup_{\substack{\alpha_{1},\alpha_{2},\beta_{2}\in\N^{n}_{0} \\ |\alpha_{1}|+|\alpha_{2}|+|\beta_{2}|\leq m}}\sup_{\substack{\theta2\in K_{2} \\ g_{1},g_{2}\in G}}\!\!\left|\left(\!R^{\alpha_{1}}_{1}R^{\alpha_{2}}_{2}\partial^{\beta_{2}}_{\theta_{2}}f\!\right)\!(\theta_{2},g_{2}g_{1})\right| \\ \nonumber
 & \leq C_{m}\!\!\!\!\!\!\!\!\!\!\sup_{\substack{\alpha_{1},\alpha_{2},\beta_{2}\in\N^{n}_{0} \\ |\alpha_{1}|+|\alpha_{2}|+|\beta_{2}|\leq m}}\sup_{\substack{\theta\in K_{2} \\ g\in G}}\!\left|\left(\!R^{\alpha_{1}+\alpha_{2}}\partial^{\beta_{2}}_{\theta}f\!\right)\!(\theta,g)\right| \\ \nonumber
 & = C_{m}||f||_{m,K_{2}},\ \textup{for some}\ C_{m}>0,
\end{align}
where $m\in\N_{0}$ and $K_{1},K_{2}\sqsubset\fg^{*}$ are compact. To arrive at the inequality in the next to last line, we used the fact that the commutator $[R_{i},R_{j}]=-f_{ij}^{k}R_{k}$ reduces the order of derivatives. The proof of the continuity of $\tilde{p}^{\uR *}_{ll'}$ is analogous: First, we reduce it to showing that the map $\tilde{p}^{\uR *}:C^{\infty}(G)\rightarrow C^{\infty}(G^{\times2}),\  (\tilde{p}^{\uR *}f)((\theta_{1},\theta_{2}),(g_{1},g_{2})):=f(\textup{Ad}^{*}_{g_{2}}(\theta_{1}),g_{2}g_{1}),$ is continuous. Second, we use that $\textup{Ad}^{*}:G\rightarrow GL(\fg^{*})$ gives orthogonal transformations w.r.t. to some $G$-invariant metric on $\fg^{*}$, and $(\theta,g)\mapsto ||\textup{Ad}^{*}_{g}(\theta)||_{\fg^{*}}$ is bounded on $G\times K$ for some compact subset $K\sqsubset\fg^{*}$.\\[0.1cm]
Finally, commutativity of the diagram \eqref{eq:edgeinversioncompatibilitydiagram} follows from:
\begin{align}
\label{eq:edgeinversioncompatibilityproof}
(\tilde{p}^{\uR}_{l^{-1}l'^{-1}}\circ\tilde{p}_{l'^{-1}l'})(\{(\theta_{e'},g_{e'})\}_{e'\in E(\gamma')}) & = \tilde{p}^{\uR}_{l^{-1}l'^{-1}}(\{(-\textup{Ad}^{*}_{g^{-1}_{e'}}(\theta_{e'}),g^{-1}_{e'})\}_{e'\in E(\gamma')}) \\ \nonumber
 & = \{(\textup{Ad}^{*}_{g^{-1}_{e'_{1_{e}}}...g^{-1}_{e'_{(m-1)_{e}}}}(-\textup{Ad}^{*}_{g_{e'_{m_{e}}}}(\theta_{e'_{m_{e}}})),g^{-1}_{e'_{1_{e}}}...g^{-1}_{e'_{m_{e}}})\}_{e\in E(\gamma)} \\ \nonumber
 & = \{(-\textup{Ad}^{*}_{(g_{e'_{m_{e}}}...g_{e'_{1_{e}}})^{-1}}(\theta_{e'_{m_{e}}}),(g_{e'_{m_{e}}}...g_{e'_{1_{e}}})^{-1})\}_{e\in E(\gamma)} \\ \nonumber
 & = \tilde{p}_{l^{-1}l}(\{(\theta_{e'_{m_{e}}},g_{e'_{m_{e}}}...g_{e'_{1_{e}}})\}_{e\in E(\gamma)}) \\ \nonumber
 & = (\tilde{p}_{l^{-1}l}\circ\tilde{p}^{\uL}_{ll'})(\{(\theta_{e'},g_{e'})\}_{e'\in E(\gamma')}).
\end{align}
\end{Proof}
\end{Theorem}
On the (quantum) level of continuous, linear operators $L(C^{\infty}(C_{l}))$ (which is associated with $C(G)\rtimes_{\uL}G^{\otimes|E(\gamma)|}$ in a natural way) acting on $C^{\infty}(C_{l})\subset L^{2}(C_{l})$, we have *-morphisms $\alpha^{\uR,\uL}_{l'l}$ corresponding to the Poisson maps $\tilde{p}^{\uR,\uL *}_{ll'}:C^{\infty}(\Gamma_{l})\rightarrow C^{\infty}(\Gamma_{l'})$\footnote{We consider only *-morphisms corresponding to $\tilde{p}^{\uR*}_{ll'}$ and $\tilde{p}^{\uL*}_{ll'}$, because generic $\tilde{p}^{c*}_{ll'}$ are not Poisson maps. Furthermore, we have compatibility with transitivity of $\leq$, and the commutative diagram \eqref{eq:edgeinversioncompatibilitydiagram} w.r.t. to edge inversion for $\tilde{p}^{\uR*}_{ll'}$ and $\tilde{p}^{\uL*}_{ll'}$.} via quantisation (Kohn-Nirenberg and Weyl, see paragraph III.A.2 of our second article\cite{StottmeisterCoherentStatesQuantumII}).
\begin{Theorem}
\label{thm:loopphasespacequant}
Given $l,l'\in\cL$, s.t. $l\lesssim l'$, we represent the continuous, linear operators in $L(C^{\infty}(C_{l}))$ and $L(C^{\infty}(C_{l'}))$ by their (left) convolution kernels (obtained from Schwartz' kernel theorem). Then, we have injective *-morphisms (*-isomorphisms for $l\simeq l'$)
\begin{align}
\label{loopphasespacequantmorph}
\alpha^{\uR,\uL}_{l'l}:\ \cD'(C_{l})\hat{\otimes}\ \!C^{\infty}(C_{l}) & \rightarrow \cD'(C_{l'})\hat{\otimes}\ \!C^{\infty}(C_{l'}),
\end{align}
induced from the four fundamental injective *-morphisms:
\begin{align}
\label{eq:algebraautsleftright}
\eta: L(C^{\infty}(G)) & \rightarrow L(C^{\infty}(G^{2})), & \eta(F)((h_{2},g_{2}),\!(h_{1},g_{1})) & := \delta_{e}(h_{1})F(h_{2},g_{2}),\\ \nonumber
\gamma: L(C^{\infty}(G)) & \rightarrow L(C^{\infty}(G)), & \gamma(F)(h,g) & := I(F)(h,g^{-1}) = F(\alpha_{g^{-1}}(h^{-1}),g^{-1}), \\ \nonumber
\alpha^{\uR}: L(C^{\infty}(G)) & \rightarrow L(C^{\infty}(G^{2})), & \alpha^{\uR}(F)((h_{2},g_{2}),\!(h_{1},g_{1})) & := \delta_{e}(h_{2})((\alpha^{*}_{g_{2}}\otimes L^{*}_{g_{2}})F)(h_{1},g_{1}\!) \\ \nonumber
&  &  & \ = \delta_{e}(h_{2})F(\alpha_{g_{2}}(h_{1}),g_{2}g_{1}\!), \\ \nonumber
\alpha^{\uL}: L(C^{\infty}(G)) & \rightarrow L(C^{\infty}(G^{2})), & \alpha^{\uL}(F)((h_{2},g_{2}),\!(h_{1},g_{1})) & := \delta_{e}(h_{1})(R^{*}_{g_{1}}F)(h_{2},g_{2}) \\ \nonumber
&  &  & \ = \delta_{e}(h_{1})F(h_{2},g_{2}g_{1}),
\end{align}
for $F\in\cD'(G)\hat{\otimes}\ \!C^{\infty}(G)$. Furthermore, we have commutative diagrams for $l\lesssim l'$:
\begin{align}
\label{eq:groupdoublingautcomdiag}
\xymatrix{
\hat{\mathcal{E}}'_{U_{l}}(\mathfrak{c}^{*}_{l})\hat{\otimes}\ \!C^{\infty}(C_{l}) \ar[r]^{\tilde{p}^{\uR,\uL*}_{ll'}} \ar[d]_{F^{(W),\varepsilon}}& \hat{\mathcal{E}}'_{U_{l'}}(\mathfrak{c}^{*}_{l'})\hat{\otimes}\ \!C^{\infty}(C_{l'}) \ar[d]^{F^{(W),\varepsilon}} \\
\cD'(C_{l})\hat{\otimes}\ \!C^{\infty}(C_{l}) \ar[r]_{\alpha^{\uR,\uL}_{l'l}} & \cD'(C_{l'})\hat{\otimes}\ \!C^{\infty}(C_{l'}),
}
\end{align}
where we used the notation of paragraph III.A.2 of our second article\cite{StottmeisterCoherentStatesQuantumII}, and $\mathfrak{c}^{*}_{l},\ l\in\cL,$ denotes the dual of the Lie algebra of $C_{l}$.\\[0.1cm]
The maps $\alpha^{\uR,\uL}_{l'l}$ respect transitivity of $\leq$, i.e. $\alpha^{\uR,\uL}_{l''l'}\circ\alpha^{\uR,\uL}_{l'l}=\alpha^{\uR,\uL}_{l''l}$ for $l\leq l'\leq l''$. \\
Also for $\lessdot_{\uR,\uL}$, we have transitivity of the corresponding collections of maps $\alpha^{\uR,\uL}_{l'l}$, i.e. $\alpha^{\uR,\uL}_{l''l'}\circ\alpha^{\uR,\uL}_{l'l}=\alpha^{\uR,\uL}_{l''l}$ for $l\lessdot_{\uR,\uL} l'\lessdot_{\uR,\uL} l''$.\\
If $\gamma\subset\gamma'$ and $l^{-1}, l'^{-1}$ denote the structured graphs with all edge orientations reversed, and \\
$\alpha_{ll^{-1}}:\cD'(C_{l^{-1}})\hat{\otimes}\ \!C^{\infty}(C_{l^{-1}})\rightarrow \cD'(C_{l})\hat{\otimes}\ \!C^{\infty}(C_{l}),\ l\in\cL,$ are the edge inversion *-isomorphisms ($\alpha^{\uR}_{ll^{-1}}=\alpha^{\uL}_{ll^{-1}}$), we have:
\begin{align}
\label{eq:edgeinversionautcompatibilitydiagram}
\xymatrix{
 \cD'(C_{l^{-1}})\hat{\otimes}\ \!C^{\infty}(C_{l^{-1}}) \ar[rr]^{\alpha_{ll^{-1}}} \ar[d]_{\alpha^{\uL}_{l'^{-1}l^{-1}}} &  &  \cD'(C_{l})\hat{\otimes}\ \!C^{\infty}(C_{l}) \ar[d]^{\alpha^{\uR}_{l'l}} \\
 \cD'(C_{l'^{-1}})\hat{\otimes}\ \!C^{\infty}(C_{l'^{-1}}) \ar[rr]_{\alpha_{l'l'^{-1}}} &  &  \cD'(C_{l'})\hat{\otimes}\ \!C^{\infty}(C_{l'})
}
\end{align}
\begin{Proof}
Since $l\lesssim l'$, we know that $|\gamma|\subseteq|\gamma'|$, i.e. $\gamma$ is obtained from $\gamma'$ by removing, inverting and composing edges. These operations are modelled by the four fundamental maps \eqref{eq:algebraautsleftright}. Therefore, we only need to understand how an operator in $L(C^{\infty}(G))$ behaves w.r.t these, and whether the prescriptions really define *-morphisms. Thus, we may reduce the proof to showing that the maps \eqref{eq:algebraautsleftright} define injective *-morphisms.\\[0.1cm]
Let us first show injectivity: The injectivity of $\gamma$ follows from the injectivity of $I$ (see proposition \ref{prop:loopphasespaceedgeinversion} and \eqref{eq:l1inversioniso}). Injectivity of $\alpha^{\uL},\ \alpha^{\uR}\ \&\ \eta$ can be deduced in the following way: Assume we are given $F,F'\in\cD'(G)\hat{\otimes}\ \!C^{\infty}(G)$ s.t. $\alpha(F)=\alpha(F')$. Then, we define $\Psi_{1}\in C^{\infty}(G^{\times2}):\ \Psi_{1}(g_{1},g_{2}):=\Psi(g_{2})$ and $\Psi_{2}\in C^{\infty}(G^{\times2}):\ \Psi_{2}(g_{1},g_{2}):=\Psi(g_{2}g_{1})$ for $\Psi\in C^{\infty}(G)$. Applying $\rho_{\uL}(\eta(F))$ and $\rho_{\uL}(\eta(F'))$ to $\Psi_{1}$, we find:
\begin{align}
\label{eq:groupdoublingautinj}
(\rho_{\uL}(\eta(F))\Psi_{1})(g_{1},g_{2})  & = \int_{G}dh_{1}\int_{G}dh_{2}\ \eta(F)((h_{2},g_{2}),(h_{1},g_{1}))\Psi_{1}(h^{-1}_{1}g_{1},h^{-1}_{2}g_{2}) \\ \nonumber
 & = \int_{G}dh_{2}\ F(h_{2},g_{2})\Psi(h^{-1}_{2}g_{2}) \\ \nonumber
 & = (\rho_{\uL}(F)\Psi)(g_{2}) \\ \nonumber
 (\rho_{\uL}(\eta(F'))\Psi_{1})(g_{1},g_{2})  & = \int_{G}dh_{1}\int_{G}dh_{2}\ \eta(F')((h_{2},g_{2}),(h_{1},g_{1}))\Psi_{1}(h^{-1}_{1}g_{1},h^{-1}_{2}g_{2}) \\ \nonumber
 & = \int_{G}dh_{2}\ F'(h_{2},g_{2})\Psi(h^{-1}_{2}g_{2}) \\ \nonumber
 & = (\rho_{\uL}(F'))\Psi)(g_{2}),
\end{align}
which shows that $\rho_{\uL}(F)=\rho_{\uL}(F')$, and therefore $F=F'$.\\
Applying $\rho_{\uL}(\alpha^{\uL}(F))$ and $\rho_{\uL}(\alpha^{\uL}(F'))$ to $\Psi_{2}$, we find:
\begin{align}
\label{eq:groupdoublingtensorautinj}
(\rho_{\uL}(\alpha^{\uL}(F))\Psi_{2})(g_{1},g_{2})  & = \int_{G}dh_{1}\int_{G}dh_{2}\ \alpha^{\uL}(F)((h_{2},g_{2}),(h_{1},g_{1}))\Psi_{2}(h^{-1}_{1}g_{1},h^{-1}_{2}g_{2}) \\ \nonumber
 & = \int_{G}dh_{2}\ F(h_{2},g_{2}g_{1})\Psi(h^{-1}_{2}g_{2}g_{1}) \\ \nonumber
 & = (\rho_{\uL}(F)\Psi)(g_{2}g_{1}) \\ \nonumber
 (\rho_{\uL}(\alpha^{\uL}(F'))\Psi_{2})(g_{1},g_{2})  & = \int_{G}dh_{1}\int_{G}dh_{2}\ \alpha^{\uL}(F')((h_{2},g_{2}),(h_{1},g_{1}))\Psi_{2}(h^{-1}_{1}g_{1},h^{-1}_{2}g_{2}) \\ \nonumber
 & = \int_{G}dh_{2}\ F'(h_{2},g_{2}g_{1})\Psi(h^{-1}_{2}g_{2}g_{1}) \\ \nonumber
 & = (\rho_{\uL}(F')\Psi)(g_{2}g_{1}),
\end{align}
which shows that $\rho_{\uL}(F)=\rho_{\uL}(F')$, and therefore $F=F'$. An analogous calculation works for $\alpha^{\uR}$.\\[0.1cm]
The *-morphism property needs to be proved w.r.t. to the involution and convolution product of $C(G)\rtimes_{\uR/\uL}G$ (see definition II.4 of our companion article\cite{StottmeisterCoherentStatesQuantumII}), because we work with (left) convolution kernels (linearity of $\eta,\ \gamma\ \&\ \alpha^{\uR,\uL}$ is evident). $\forall F,F'\in\cD'(G)\hat{\otimes}\ \!C^{\infty}(G)$: 
\begin{align}
\label{eq:groupdoublingautleftmorph}
\alpha^{\uL}(F\!\ast_{\uL}\!F')((h_{2},g_{2}),(h_{1},g_{1})) & = \delta_{e}(h_{1})(F\!\ast_{\uL}\!F')(h_{2},g_{2}g_{1}) \\ \nonumber
 & = \delta_{e}(h_{1})\int_{G}dh\ F(h,g_{2}g_{1})F'(h^{-1}h_{2},h^{-1}g_{2}g_{1}) \\ \nonumber
 & = \int_{G}dh'\int_{G}dh\ \delta_{e}(h')F(h,g_{2}g_{1})\delta_{e}(h^{-1}h_{1})F'(h^{-1}h_{2},h^{-1}g_{2}h'^{-1}g_{1}) \\ \nonumber
 & = \int_{G}dh'\int_{G}dh\ \alpha^{\uL}(F)((h,g_{2}),(h',g_{1})) \\[-0.25cm] \nonumber
 & \hspace{2.5cm}\times\alpha^{\uL}(F')((h^{-1}h_{2},h^{-1}g_{2}),(h'^{-1}h_{1},h'^{-1}g_{1})) \\ \nonumber
 & = (\alpha^{\uL}(F)\!\ast_{\uL}\!\alpha^{\uL}(F'))((h_{2},g_{2}),(h_{1},g_{1})). \\[0.25cm] \nonumber
\alpha^{\uL}(F^{*_{\uL}})((h_{2},g_{2}),(h_{1},g_{1})) & = \delta_{e}(h_{1})F^{*_{\uL}}(h_{2},g_{2}g_{1}) \\ \nonumber
 & = \delta_{e}(h_{1})\overline{F(h^{-1}_{2},h^{-1}_{2}g_{2}g_{1})} \\ \nonumber
 & = \overline{\delta_{e}(h^{-1}_{1})F(h^{-1}_{2},h^{-1}_{2}g_{2}h^{-1}_{1}g_{1})} \\ \nonumber
 & = \overline{\alpha^{\uL}(F)((h^{-1}_{2},h^{-1}_{2}g_{2}),(h^{-1}_{1},h^{-1}_{1}g_{1}))} \\ \nonumber
 & = \alpha^{\uL}(F)^{*_{\uL}}((h_{2},g_{2}),(h_{1},g_{1})),
\\
\label{eq:groupdoublingautrightmorph}
\alpha^{\uR}(F\!\ast_{\uL}\!F')((h_{2},g_{2}),\!(h_{1},g_{1})) & \!=\! \delta_{e}(h_{2})(F\ast_{\uL}\!F')(\alpha_{g_{2}}(h_{1}),g_{2}g_{1}) \\ \nonumber
 & = \delta_{e}(h_{2})\int_{G}\!\!dh\ \!F(h,g_{2}g_{1})F'(h^{-1}\!\alpha_{g_{2}}(h_{1}),h^{-1}\!g_{2}g_{1}) \\ \nonumber
 & = \int_{G}\!\!dh'\!\!\int_{G}\!\!dh\ \!\delta_{e}(h')F(h,g_{2}g_{1})\delta_{e}(h'^{-1}\!h_{2})F'(h^{-1}\!\alpha_{g_{2}}(h_{1}),h'^{-1}\!h^{-1}\!g_{2}g_{1}) \\ \nonumber
 & = \int_{G}\!\!dh'\!\!\int_{G}\!\!dh\ \!\delta_{e}(h')F(\alpha_{h'^{-1}\!g_{2}}(h),g_{2}g_{1})\delta_{e}(h'^{-1}\!h_{2}) \\[-0.25cm] \nonumber
 & \hspace{2.5cm}\times F'(\alpha_{h'^{-1}\!g_{2}}(h^{-1}\!h_{1}),h'^{-1}\!g_{2}h^{-1}\!g_{1}) \\ \nonumber
 & = \int_{G}\!\!dh'\!\!\int_{G}\!\!dh\ \!\alpha^{\uR}(F)((h',g_{2}),(h,g_{1})) \\[-0.25cm] \nonumber
 & \hspace{2.5cm}\times\alpha^{\uR}(F')((h'^{-1}\!h_{2},h'^{-1}\!g_{2}),(h^{-1}\!h_{1},h^{-1}\!g_{1})) \\ \nonumber
 & = (\alpha^{\uR}(F)\!\ast_{\uL}\!\alpha^{\uR}(F'))((h_{2},g_{2}),(h_{1},g_{1})). \\[0.25cm] \nonumber
\alpha^{\uR}(F^{*_{\uL}})((h_{2},g_{2}),(h_{1},g_{1})) & \!=\! \delta_{e}(h_{2})F^{*_{\uL}}(\alpha_{g_{2}}(h_{1}),g_{2}g_{1}) \\ \nonumber
 & = \delta_{e}(h_{2})\overline{F(\alpha_{g_{2}}(h_{1})^{-1},\alpha_{g_{2}}(h_{1})^{-1}g_{2}g_{1})} \\ \nonumber
 & = \overline{\delta_{e}(h^{-1}_{2})F(\alpha_{h^{-1}_{2}g_{2}}(h_{1})^{-1},h^{-1}_{2}g_{2}h_{1}^{-1}g_{1})} \\ \nonumber
 & = \overline{\alpha^{\uR}(F)((h^{-1}_{2},h^{-1}_{2}g_{2}),(h^{-1}_{1},h^{-1}_{1}g_{1}))} \\ \nonumber
 & = \alpha^{\uR}(F)^{*_{\uL}}((h_{2},g_{2}),(h_{1},g_{1})),
\\
\label{eq:groupremovalautmorph}
\eta(F\ast_{\uL}F')((h_{2},g_{2}),(h_{1},g_{1})) & = \delta_{e}(h_{1})(F\ast_{\uL}F')(h_{2},g_{2}) \\ \nonumber
 & = \delta_{e}(h_{1})\int_{G}dh\ F(h,g_{2})F'(h^{-1}h_{2},h^{-1}g_{2}) \\ \nonumber
 & = \int_{G}dh'\int_{G}dh\ \delta_{e}(h')F(h,g_{2})\delta_{e}(h'^{-1}h_{1})F'(h^{-1}h_{2},h^{-1}g_{2}) \\ \nonumber
 & = \int_{G}dh'\int_{G}dh\ \eta(F)((h,g_{2}),(h',g_{1})) \\[-0.25cm] \nonumber
 & \hspace{2.5cm}\times\eta(F')((h^{-1}h_{2},h^{-1}g_{2}),(h'^{-1}h_{1},h'^{-1}g_{1})) \\ \nonumber
 & = (\eta(F)\ast_{\uL}\eta(F'))((h_{2},g_{2}),(h_{1},g_{1})). \\[0.25cm] \nonumber
\eta(F^{*_{\uL}})((h_{2},g_{2}),(h_{1},g_{1})) & = \delta_{e}(h_{1})F^{*_{\uL}}(h_{2},g_{2}) \\ \nonumber
 & = \delta_{e}(h_{1})\overline{F(h^{-1}_{2},h^{-1}_{2}g_{2})} \\ \nonumber
 & = \overline{\delta_{e}(h^{-1}_{1})F(h^{-1}_{2},h^{-1}_{2}g_{2})} \\ \nonumber
 & = \overline{\eta(F)((h^{-1}_{2},h^{-1}_{2}g_{2}),(h^{-1}_{1},h^{-1}_{1}g_{1}))} \\ \nonumber
 & = \eta(F)^{*_{\uL}}((h_{2},g_{2}),(h_{1},g_{1})),\\
\label{eq:groupinversionautmorph}
\gamma(F\!\ast_{\uL}\!F')(h,g) & = I(F\!\ast_{\uL}\!F')(h,g^{-1}) \\ \nonumber
 & = \int_{G}\!\!dk\ \!F(k,g^{-1})F'(k^{-1}\alpha_{g^{-1}}(h^{-1}),k^{-1}g^{-1}) \\ \nonumber
 & = \int_{G}\!\!dk\ \!F(\alpha_{g^{-1}}(k),g^{-1})F'(\alpha_{g^{-1}}(hk)^{-1},(kg)^{-1}) \\ \nonumber
 & = \int_{G}\!\!dk\ \!F(\alpha_{g^{-1}}(k^{-1}),g^{-1})F'(\alpha_{g^{-1}k}(k^{-1}h)^{-1},(k^{-1}g)^{-1}) \\ \nonumber
 & = \int_{G}\!\!dk\ \!\gamma(F)(k,g)\gamma(F')(k^{-1}h,k^{-1}g) \\ \nonumber
 & = (\gamma(F)\ast_{\uL}\!\gamma(F'))(h,g). \\[0.25cm] \nonumber
\gamma(F^{*_{\uL}})(h,g) & = F^{*_{\uL}}(\alpha_{g^{-1}}(h^{-1}),g^{-1}) \\ \nonumber
 & = \overline{F(\alpha_{g^{-1}}(h),g^{-1}h)} \\ \nonumber
 & = \overline{\gamma(F)(h^{-1},h^{-1}g)} \\ \nonumber
 & = \gamma(F)^{*_{\uL}}(h,g).
\end{align}
Now, let $\sigma\in\hat{\mathcal{E}}'_{U}(\fg^{*})\hat{\otimes}\ \!C^{\infty}(G)$.Then, $\tilde{p}^{\uL*}\sigma\in\hat{\mathcal{E}}'_{U\times U}((\fg^{*})^{\times2})\hat{\otimes}\ \!C^{\infty}(G^{\times2})$, because $(\theta_{1},\theta_{2})\mapsto\sigma(\theta_{2},g_{2}g_{1})$ is analytic for any $g_{1},g_{2}\in G$, with constant growth bound in $\theta_{1}$, and $(g_{1},g_{2})\mapsto\sigma(\theta_{2},g_{2}g_{1})$ is smooth for any $\theta_{2}\in\fg^{*}$. Similarly, $\tilde{p}^{\uR*}\sigma,\ \tilde{r}^{*}\sigma\in\hat{\mathcal{E}}'_{U\times U}((\fg^{*})^{\times2})\hat{\otimes}\ \!C^{\infty}(G^{\times2})$ and $\tilde{\iota}^{*}\sigma\in\hat{\mathcal{E}}'_{U}(\fg^{*})\hat{\otimes}\ \!C^{\infty}(G)$ (the coadjoint action, $\textup{Ad}^{*}$, is analytic). Finally, we observe:
\begin{align}
\label{eq:groupdoublingautleft}
F^{W,\varepsilon}_{\tilde{p}^{\uL*}\sigma}((h_{2},g_{2}),(h_{1},g_{1})) & = \int_{\fg^{*}}\frac{d\theta_{1}}{(2\pi\varepsilon)^{n}}\int_{\fg^{*}}\frac{d\theta_{2}}{(2\pi\varepsilon)^{n}}e^{\frac{i}{\varepsilon}(\theta_{1}(X_{h_{1}})+\theta_{2}(X_{h_{2}}))}\\[-0.25cm] \nonumber
 & \hspace{3.5cm}\times(\tilde{p}^{\uL*}\sigma)((\theta_{2},\sqrt{h^{-1}_{2}}g_{2}),(\theta_{1},\sqrt{h^{-1}_{1}}g_{1})) \\ \nonumber
 & = \int_{\fg^{*}}\frac{d\theta_{1}}{(2\pi\varepsilon)^{n}}\int_{\fg^{*}}\frac{d\theta_{2}}{(2\pi\varepsilon)^{n}}e^{\frac{i}{\varepsilon}(\theta_{1}(X_{h_{1}})+\theta_{2}(X_{h_{2}}))}\sigma(\theta_{2},\sqrt{h^{-1}_{2}}g_{2}\sqrt{h^{-1}_{1}}g_{1}) \\ \nonumber
 & = \delta^{(n)}_{0}(X_{h_{1}})\int_{\fg^{*}}\frac{d\theta_{2}}{(2\pi\varepsilon)^{n}}e^{\frac{i}{\varepsilon}\theta_{2}(X_{h_{2}})}\sigma(\theta_{2},\sqrt{h^{-1}_{2}}g_{2}\sqrt{h^{-1}_{1}}g_{1}) \\ \nonumber
 &\hspace{-0.29cm} \underset{j(0)=1}{=} \delta_{e}(h_{1})\int_{\fg^{*}}\frac{d\theta_{2}}{(2\pi\varepsilon)^{n}}e^{\frac{i}{\varepsilon}\theta_{2}(X_{h_{2}})}\sigma(\theta_{2},\sqrt{h^{-1}_{2}}g_{2}g_{1}) \\ \nonumber
 & = \delta_{e}(h_{1})F^{W,\varepsilon}_{\sigma}(h_{2},g_{2}g_{1}) \\ \nonumber
 & = \alpha^{\uL}(F^{W,\varepsilon}_{\sigma})((h_{2},g_{2}),(h_{1},g_{1})),
\\[0.5cm]
\label{eq:groupdoublingautright}
F^{W,\varepsilon}_{\tilde{p}^{\uR*}\sigma}((h_{2},g_{2}),(h_{1},g_{1})) & = \int_{\fg^{*}}\frac{d\theta_{1}}{(2\pi\varepsilon)^{n}}\int_{\fg^{*}}\frac{d\theta_{2}}{(2\pi\varepsilon)^{n}}e^{\frac{i}{\varepsilon}(\theta_{1}(X_{h_{1}})+\theta_{2}(X_{h_{2}}))}\\[-0.25cm] \nonumber
 & \hspace{3.5cm}\times(\tilde{p}^{\uR*}\sigma)((\theta_{2},\sqrt{h^{-1}_{2}}g_{2}),(\theta_{1},\sqrt{h^{-1}_{1}}g_{1})) \\ \nonumber
 & = \int_{\fg^{*}}\frac{d\theta_{1}}{(2\pi\varepsilon)^{n}}\int_{\fg^{*}}\frac{d\theta_{2}}{(2\pi\varepsilon)^{n}}e^{\frac{i}{\varepsilon}(\theta_{1}(X_{h_{1}})+\theta_{2}(X_{h_{2}}))}\sigma(\textup{Ad}^{*}_{g_{2}}(\theta_{1}),\sqrt{h^{-1}_{2}}g_{2}\sqrt{h^{-1}_{1}}g_{1}) \\ \nonumber
 & = \delta^{(n)}_{0}(X_{h_{2}})\int_{\fg^{*}}\frac{d\theta_{1}}{(2\pi\varepsilon)^{n}}e^{\frac{i}{\varepsilon}\textup{Ad}^{*}_{g^{-1}_{2}}(\theta_{1})(X_{h_{1}})}\sigma(\theta_{1},\sqrt{h^{-1}_{2}}g_{2}\sqrt{h^{-1}_{1}}g_{1}) \\ \nonumber
 &\hspace{-0.29cm} \underset{j(0)=1}{=} \delta_{e}(h_{2})\int_{\fg^{*}}\frac{d\theta_{1}}{(2\pi\varepsilon)^{n}}e^{\frac{i}{\varepsilon}\theta_{1}(X_{\alpha_{g_{2}}(h_{1})})}\sigma(\theta_{1},g_{2}\sqrt{h^{-1}_{1}}g^{-1}_{2}g_{2}g_{1}) \\ \nonumber
 &\hspace{-1.55cm} \underset{g_{2}\sqrt{h^{-1}_{1}}g^{-1}_{2}=\sqrt{\alpha_{g_{2}}(h_{1})^{-1}}}{=} \delta_{e}(h_{2})F^{W,\varepsilon}_{\sigma}(\alpha_{g_{2}}(h_{1}),g_{2}g_{1}) \\ \nonumber
 & = \alpha^{\uR}(F^{W,\varepsilon}_{\sigma})((h_{2},g_{2}),(h_{1},g_{1})),
\\[0.5cm]
\label{eq:groupremovalaut}
F^{W,\varepsilon}_{\tilde{r}^{*}\sigma}((h_{2},g_{2}),(h_{1},g_{1})) & = \int_{\fg^{*}}\frac{d\theta_{1}}{(2\pi\varepsilon)^{n}}\int_{\fg^{*}}\frac{d\theta_{2}}{(2\pi\varepsilon)^{n}}e^{\frac{i}{\varepsilon}(\theta_{1}(X_{h_{1}})+\theta_{2}(X_{h_{2}}))}\\[-0.25cm] \nonumber
 & \hspace{3.5cm}\times(\tilde{r}^{*}\sigma)((\theta_{2},\sqrt{h^{-1}_{2}}g_{2}),(\theta_{1},\sqrt{h^{-1}_{1}}g_{1})) \\ \nonumber
 & = \int_{\fg^{*}}\frac{d\theta_{1}}{(2\pi\varepsilon)^{n}}\int_{\fg^{*}}\frac{d\theta_{2}}{(2\pi\varepsilon)^{n}}e^{\frac{i}{\varepsilon}(\theta_{1}(X_{h_{1}})+\theta_{2}(X_{h_{2}}))}\sigma(\theta_{2},\sqrt{h^{-1}_{2}}g_{2}) \\ \nonumber
 & = \delta^{(n)}_{0}(X_{h_{1}})\int_{\fg^{*}}\frac{d\theta_{2}}{(2\pi\varepsilon)^{n}}e^{\frac{i}{\varepsilon}\theta_{2}(X_{h_{2}})}\sigma(\theta_{2},\sqrt{h^{-1}_{2}}g_{2}) \\ \nonumber
 &\hspace{-0.29cm} \underset{j(0)=1}{=} \delta_{e}(h_{1})\int_{\fg^{*}}\frac{d\theta_{2}}{(2\pi\varepsilon)^{n}}e^{\frac{i}{\varepsilon}\theta_{2}(X_{h_{2}})}\sigma(\theta_{2},\sqrt{h^{-1}_{2}}g_{2}) \\ \nonumber
 & = \eta(F^{W,\varepsilon}_{\sigma})((h_{2},g_{2}),(h_{1},g_{1})),
\\[0.5cm]
\label{eq:groupinversionaut}
F^{W,\varepsilon}_{\tilde{\iota}^{*}\sigma}(h,g) & = \int_{\fg^{*}}\frac{d\theta}{(2\pi\varepsilon)^{n}}e^{\frac{i}{\varepsilon}\theta(X_{h})}(\tilde{\iota}^{*}\sigma)(\theta,\sqrt{h^{-1}}g) \\ \nonumber
 & = \int_{\fg^{*}}\frac{d\theta}{(2\pi\varepsilon)^{n}}e^{\frac{i}{\varepsilon}\theta(X_{h})}\sigma(-\textup{Ad}^{*}_{g^{-1}\sqrt{h}}(\theta),g^{-1}\sqrt{h}) \\ \nonumber
 & = \int_{\fg^{*}}\frac{d\theta}{(2\pi\varepsilon)^{n}}e^{-\frac{i}{\varepsilon}\textup{Ad}^{*}_{\sqrt{h^{-1}}g}(\theta)(X_{h})}\sigma(\theta,g^{-1}\sqrt{h}) \\ \nonumber
 & = \int_{\fg^{*}}\frac{d\theta}{(2\pi\varepsilon)^{n}}e^{\frac{i}{\varepsilon}\theta(X_{\alpha_{g^{-1}\sqrt{h}}(h^{-1})})}\sigma(\theta,g^{-1}\sqrt{h}) \\ \nonumber
 & = \int_{\fg^{*}}\frac{d\theta}{(2\pi\varepsilon)^{n}}e^{\frac{i}{\varepsilon}\theta(X_{\alpha_{g^{-1}}(h^{-1})})}\sigma(\theta,g^{-1}\sqrt{h}gg^{-1}) \\ \nonumber
 & = F^{W,\varepsilon}_{\sigma}(\alpha_{g^{-1}}(h^{-1}),g^{-1}) \\ \nonumber
 & = \gamma(F^{W,\varepsilon}_{\sigma})(h,g),
\end{align}
which proves \eqref{eq:groupdoublingautcomdiag} for Weyl quantisation. The proof for the Kohn-Nirenberg quantisation is analogous.\\[0.1cm]
To show the transitivity property, we argue in the same fashion as in the proof of theorem \ref{thm:loopphasespacepartialordercompatible}. First, we analyse the maps (cp. \eqref{eq:projectiontransitivity}):
\begin{align}
\label{eq:injectiontransitivity}
& \alpha^{\uL}_{2'1}: \cD'(G)\hat{\otimes}\ \!C^{\infty}(G)) \rightarrow \cD'(G_{2'}\times G_{1})\hat{\otimes}\ \!C^{\infty}(G_{2'}\times G_{1})), \\ \nonumber
& \alpha^{\uL}_{2'1}(F)((h_{2'},g_{2'}),(h_{1},g_{1})) = \delta_{e}(h_{1})F(h_{2'},g_{2'}g_{1}), \\ \nonumber
& \alpha^{\uL}_{(32)1}: \cD'(G_{2'}\times G_{1})\hat{\otimes}\ \!C^{\infty}(G_{2'}\times G_{1})) \rightarrow \cD'(G_{3}\times G_{2}\times G_{1})\hat{\otimes}\ \!C^{\infty}(G_{3}\times G_{2}\times G_{1})), \\ \nonumber
& \alpha^{\uL}_{(32)1}(F)((h_{3},g_{3}),(h_{2},g_{2}),(h_{1},g_{1})) = \delta_{e}(h_{2})F((h_{3},g_{3}g_{2}),(h_{1},g_{1})), \\[0.1cm] \nonumber
& \alpha^{\uL}_{31'}: \cD'(G)\hat{\otimes}\ \!C^{\infty}(G)) \rightarrow \cD'(G_{3}\times G_{1'})\hat{\otimes}\ \!C^{\infty}(G_{3}\times G_{1'})), \\ \nonumber
& \alpha^{\uL}_{31'}(F)((h_{3},g_{3}),(h_{1'},g_{1'})) = \delta_{e}(h_{1'})F(h_{3},g_{3}g_{1'}), \\ \nonumber
& \alpha^{\uL}_{3(21)}: \cD'(G_{3}\times G_{1'})\hat{\otimes}\ \!C^{\infty}(G_{3}\times G_{1'})) \rightarrow \cD'(G_{3}\times G_{2}\times G_{1})\hat{\otimes}\ \!C^{\infty}(G_{3}\times G_{2}\times G_{1})), \\ \nonumber
& \alpha^{\uL}_{3(21)}(F)((h_{3},g_{3}),(h_{2},g_{2}),(h_{1},g_{1})) = \delta_{e}(h_{1})F((h_{3},g_{3}),(h_{2},g_{2}g_{1})), \\[0.1cm] \nonumber
& \alpha^{\uL}_{321}: \cD'(G)\hat{\otimes}\ \!C^{\infty}(G)) \rightarrow \cD'(G_{3}\times G_{2}\times G_{1})\hat{\otimes}\ \!C^{\infty}(G_{3}\times G_{2}\times G_{1})), \\ \nonumber
& \alpha^{\uL}_{321}(F)((h_{3},g_{3}),(h_{2},g_{2}),(h_{1},g_{1})) = \delta_{e}(h_{1})\delta_{e}(h_{2})F(h_{3},g_{3}g_{2}g_{1}) \\[0.25cm] \nonumber
\Rightarrow\ &\ (\alpha^{\uL}_{(32)1}\circ\alpha^{\uL}_{2'1})(F) = (\alpha^{\uL}_{3(21)}\circ\alpha^{\uL}_{31'})(F) = \alpha^{\uL}_{321}(F).
\end{align}
From this we understand, that the $\alpha^{\uL}_{l'l}$ embed the $h$-dependence of an edge splitting into the outmost tensor factor corresponding to the final edge in a composition chain. Since this property is preserved under successive splittings and adding of new edges to a composition chain (this is capture by $\eta$), we obtain transitivity of the $\alpha^{\uL}_{l'l}$ w.r.t. $\leq$. A similar argument works for the maps $\alpha^{\uR}_{l'l},\ l\leq l',$ as in this case the embedding, which arises from edge splitting, is into the outmost tensor factor corresponding to the initial edge in a composition chain. As before, the argument also works for $\lessdot_{\uR}$ and $\lessdot_{\uL}$ in combination with $\alpha^{\uR}_{l'l}$ and $\alpha^{\uL}_{l'l}$ respectively, because these partial orders preserve the notion of first respectively last part of an edge between an oriented graph and its oriented subgraphs (cf. \cite{LaneryProjectiveLoopQuantum}, p. 52-53). If we take edge inversion into account, the situation will change, because the notion of initial and finial edge in a composition chain gets permuted.  This is essentially captured in the diagram \eqref{eq:edgeinversionautcompatibilitydiagram}, which follows from: $\forall F\in\cD'(C_{l})\hat{\otimes}\ \!C^{\infty}(C_{l})$:
\begin{align}
\label{eq:edgeinversionautcompatibilitydiagramproof}
(\alpha^{\uR}_{l'l}\circ\alpha_{ll^{-1}})(F)\left(\{(h_{e'},g_{e'})\}_{e'\in E(\gamma')}\right) & = \Big(\prod_{e\in E(\gamma)}\delta_{e}(h_{e'_{2_{e}}})...\delta_{e}(h_{e'_{m_{e}}})\Big) \\[-0.2cm] \nonumber
 &\hspace{1.25cm} \times\alpha_{ll^{-1}}(F)\big(\{(\alpha_{g_{e'_{m_{e}}}...g_{e'_{2_{e}}}}(h_{e'_{1_{e}}}),g_{e'_{m_{e}}}...g_{e'_{1_{e}}})\}_{e\in E(\gamma)}\big) \\ \nonumber
 & = \Big(\prod_{e\in E(\gamma)}\delta_{e}(h_{e'_{2_{e}}})...\delta_{e}(h_{e'_{m_{e}}})\Big)\\[-0.2cm] \nonumber
 &\hspace{1.25cm} \times F\big(\{(\alpha_{g^{-1}_{e'_{1_{e}}}}(h^{-1}_{e'_{1_{e}}}),g^{-1}_{e'_{1_{e}}}...g^{-1}_{e'_{m_{e}}})\}_{e\in E(\gamma)}\big) \\ \nonumber
 & = \Big(\prod_{e\in E(\gamma)}\delta_{e}(\alpha_{g^{-1}_{e'_{2_{e}}}}(h^{-1}_{e'_{2_{e}}}))...\delta_{e}(\alpha_{g^{-1}_{e'_{m_{e}}}}(h^{-1}_{e'_{m_{e}}}))\Big) \\[-0.2cm] \nonumber
 &\hspace{1.25cm} \times F\big(\{(\alpha_{g^{-1}_{e'_{1_{e}}}}(h^{-1}_{e'_{1_{e}}}),g^{-1}_{e'_{1_{e}}}...g^{-1}_{e'_{m_{e}}})\}_{e\in E(\gamma)}\big) \\ \nonumber
 & = \alpha^{\uL}_{l'^{-1}l^{-1}}(F)\big(\{(\alpha_{g^{-1}_{e'}}(h^{-1}_{e'}),g^{-1}_{e'})\}_{e'\in E(\gamma')}\big) \\ \nonumber
 & = (\alpha_{l'l'^{-1}}\circ\alpha^{\uL}_{l'^{-1}l^{-1}})(F)(\{(h_{e'},g_{e'})\}_{e'\in E(\gamma')}).
\end{align}
\end{Proof}
\end{Theorem}
The next corollary explains how the system of Hilbert spaces $\{L^{2}(C_{l})\}_{l\in\cL}$ fits into the picture. \\
On the one hand, the inductive limit of $\{L^{2}(C_{l})\}_{l\in\cL}$ w.r.t. $\lesssim$ is the Ashtekar-Isham-Lewandowski Hilbert space, i.e. $\varinjlim_{(\cL,\lesssim)}L^{2}(C_{l})=L^{2}(\overline{\mathcal{A}})$.\\
The same holds for the inductive limits w.r.t. $\lessdot_{\uL}$ and $\lessdot_{\uR}$
\begin{align}
\label{eq:inductivelimitleftright}
\varinjlim_{(\cL,\lessdot_{\uL,\uR})}L^{2}(C_{l})=L^{2}(\overline{\mathcal{A}}),
\end{align}
although the edge inversion is not explicitly implemented. This can be inferred from the following argument:\\[0.1cm]
If we consider the structured graphs $l,l^{-1}$ obtained from a single edge $e:[0,1]\rightarrow\Sigma$, i.e. $\gamma=e$ and $\gamma^{-1}=e^{-1}$, we can find a decomposition of $e$ into edges $e_{1}, e_{2}$, s.t. $e=e_{2}\circ e^{-1}_{1}$. Therefore, if we construct a structured graph $l'$ from the edges $e_{1},e_{2}$, we will have $l\lessdot_{\uL}l'$ and $l^{-1}\lessdot_{\uL}l'$, as well as $l\lessdot_{\uR}(l')^{-1}$ and $l^{-1}\lessdot_{\uR}(l')^{-1}$. But then, we have for $\Psi\in L^{2}(G)$:
\begin{align}
\label{eq:leftrightequivimplementsinversion}
(p^{*}_{ll'}\Psi)(g_{2},g_{1}) & = \Psi(g_{2}g^{-1}_{1}) \\ \nonumber
 & = (\iota^{*}\Psi)(g_{1}g^{-1}_{2}) \\ \nonumber
 & = (p^{*}_{l^{-1}l'}(\iota^{*}\Psi))(g_{2},g_{1}), \\[0.25cm]
(p^{*}_{l^{-1}(l')^{-1}}\Psi)(g_{1},g_{2}) & = \Psi(g_{1}g^{-1}_{2}) \\ \nonumber
 & = (\iota^{*}\Psi)(g_{2}g^{-1}_{1}) \\ \nonumber
 & = (p^{*}_{l(l')^{-1}}(\iota^{*}\Psi))(g_{1},g_{2}).
\end{align}
Thus, the edge inversion $\iota^{*}:L^{2}(G)\rightarrow L^{2}(G),\ (\iota^{*}\Psi)(g) := \Psi(g^{-1}),$ is automatically enforced as a symmetry in the limit \eqref{eq:inductivelimitleftright}, i.e. $p^{*}_{ll'}(L^{2}(C_{l})) = p_{l^{-1}l'}(\iota^{*}(L^{2}(C_{l^{-1}})))$ and $p^{*}_{l^{-1}(l')^{-1}}(L^{2}(C_{l^{-1}})) = p_{l(l')^{-1}}(\iota^{*}(L^{2}(C_{l})))$. This justifies, why we do not differentiate between $\lessdot_{\uL}$ and $\lessdot_{\uR}$ on the Hilbert space level. \\
Moreover, the action of Weyl (and Kohn-Nirenberg) quantisation $F^{W,\varepsilon}$ via $\rho_{\uL}$ on the scale of Hilbert spaces $\{L^{2}(C_{l})\}_{l\in\cL}$ is compatible with this additional symmetry (cp. diagram \eqref{eq:loopphasespaceactioncovariancerefined}). Explicitly, we have for $\sigma\in\hat{\mathcal{E}}'_{U}(\fg^{*})\hat{\otimes}\ \!C^{\infty}(G)$ and $\Psi\in C^{\infty}(G)$ w.r.t. $\lessdot_{\uL}$:
\begin{align}
\label{eq:leftequivquantimgcomp}
\left(Q^{W}_{\varepsilon}(\tilde{p}^{\uL*}_{ll'}\sigma)(p^{*}_{ll'}\Psi)\right)(g_{2},g_{1}) & = \int_{G}dh_{2}\int_{G}dh_{1}\ \!F^{W,\varepsilon}_{\tilde{p}^{\uL*}_{ll'}\sigma}((h_{2},g_{2}),(h_{1},g_{1}))(p^{*}_{ll'}\Psi)(h^{-1}_{2}g_{2},h^{-1}_{1}g_{1}) \\ \nonumber
 & = \int_{G}dh_{2}\int_{G}dh_{1}\ \!\alpha^{\uL}_{l'l}(F^{W,\varepsilon}_{\sigma})((h_{2},g_{2}),(h_{1},g_{1}))\Psi(h^{-1}_{2}g_{2}g^{-1}_{1}h_{1}) \\ \nonumber
 & = \int_{G}dh_{2}\int_{G}dh_{1}\ \!\delta_{e}(h_{1})F^{W,\varepsilon}_{\sigma}(h_{2},g_{2}g^{-1}_{1})\Psi(h^{-1}_{2}g_{2}g^{-1}_{1}h_{1}) \\ \nonumber
 & = \int_{G}dh_{2}\ \!F^{W,\varepsilon}_{\sigma}(h_{2},g_{2}g^{-1}_{1})\Psi(h^{-1}_{2}g_{2}g^{-1}_{1}) \\ \nonumber
 & = (p^{*}_{ll'}(Q^{W}_{\varepsilon}(\sigma)\Psi))(g_{2},g_{1}) \\ \nonumber
 & = \int_{G}dh_{1}\ \!F^{W,\varepsilon}_{\sigma}(h_{1},g_{2}g^{-1}_{1})\Psi(h^{-1}_{1}g_{2}g^{-1}_{1}) \\ \nonumber
 & = \int_{G}dh_{1}\ \!F^{W,\varepsilon}_{\sigma}(\alpha_{g_{2}g^{-1}_{1}}(h^{-1}_{1}),g_{2}g^{-1}_{1})\Psi(\alpha_{g_{2}g^{-1}_{1}}(h^{-1}_{1})g_{2}g^{-1}_{1}) \\ \nonumber
 & = \int_{G}dh_{2}\int_{G}dh_{1}\ \!\delta_{e}(h_{2})F^{W,\varepsilon}_{\sigma}(\alpha_{g_{2}g^{-1}_{1}}(h^{-1}_{1}),g_{2}g^{-1}_{1})\Psi(h^{-1}_{2}g_{2}g^{-1}_{1}h_{1}) \\ \nonumber
 & = \int_{G}dh_{2}\int_{G}dh_{1}\ \!\delta_{e}(h_{2})\gamma(F^{W,\varepsilon}_{\sigma})(h_{1},g_{1}g^{-1}_{2})\Psi(h^{-1}_{2}g_{2}g^{-1}_{1}h_{1}) \\ \nonumber
 & = \int_{G}dh_{2}\int_{G}dh_{1}\ \!\delta_{e}(h_{2})F^{W,\varepsilon}_{\tilde{\iota}^{*}\sigma}(h_{1},g_{1}g^{-1}_{2})\Psi(h^{-1}_{2}g_{2}g^{-1}_{1}h_{1}) \\ \nonumber
 & = \int_{G}dh_{2}\int_{G}dh_{1}\ \!\alpha^{\uL}_{l'l^{-1}}(F^{W,\varepsilon}_{\tilde{\iota}^{*}\sigma})((h_{2},g_{2}),(h_{1},g_{1}))(\iota^{*}\Psi)(h^{-1}_{1}g_{1}g^{-1}_{2}h_{2}) \\ \nonumber
 & = \int_{G}dh_{2}\int_{G}dh_{1}\ \!F^{W,\varepsilon}_{\tilde{p}^{\uL*}_{l^{-1}l'}\tilde{\iota}^{*}\sigma}((h_{2},g_{2}),(h_{1},g_{1}))\\ \nonumber
  &\hspace{3cm} \times (p^{*}_{l^{-1}l'}\iota^{*}\Psi)((h_{2},g_{2}),(h_{1},g_{1})) \\ \nonumber
 & = \left(Q^{W}_{\varepsilon}(\tilde{p}^{\uL*}_{l^{-1}l'}\tilde{\iota}^{*}\sigma)(p^{*}_{l^{-1}l'}\iota^{*}\Psi)\right)(g_{2},g_{1}),
\end{align}
where we used the invariance properties of the Haar measure on $G$. Similarly, we have w.r.t. $\lessdot_{\uR}$:
\begin{align}
\label{eq:rightequivquantimgcomp}
\Big(Q^{W}_{\varepsilon}(\tilde{p}^{\uR*}_{l^{-1}(l')^{-1}}\tilde{\iota}^{*}\sigma)(p^{*}_{l^{-1}(l')^{-1}}\Psi)\Big)(g_{1},g_{2}) & = \Big(p^{*}_{l^{-1}(l')^{-1}}(Q^{W}_{\varepsilon}(\sigma)\Psi)\Big)(g_{1},g_{2}) \\ \nonumber
 & = \Big(Q^{W}_{\varepsilon}(\tilde{p}^{\uR*}_{l(l')^{-1}}\tilde{\iota}^{*}\sigma)(p^{*}_{l(l')^{-1}}\iota^{*}\Psi)\Big)(g_{1},g_{2}).
\end{align}
On the other hand, the would-be inductive limit
\begin{align}
\label{eq:inductivelimorientedconnections}
\varinjlim_{(\cL,\leq)}L^{2}(C_{l})=L^{2}(\overline{\mathcal{A}}^{\uparrow})
\end{align}
of $\{L^{2}(C_{l})\}_{l\in\cL}$ w.r.t. $\leq$ might give rise to a Hilbert space on ``oriented'' generalised connections $\overline{\mathcal{A}}^{\uparrow}$ (edge inversion is not necessarily a symmetry). We refer to \eqref{eq:inductivelimorientedconnections} as a would-be inductive limit, because $(\cL,\leq)$ is not directed, and thus the limit is does not necessarily exist in the category of Hilbert spaces. In analogy with the $C^{*}$-algebraic construction of $\overline{\mathcal{A}}$, it could be possible to realise $\overline{\mathcal{A}}^{\uparrow}$ as the spectrum of the $C^{*}$-closure of \eqref{eq:inductivelimorientedconnections} (in the inductive sup-norms), if the limit existed.
\begin{Corollary}
\label{cor:loophilbertspacequant}
For $l\in\cL$, the action of $L(C^{\infty}(C_{l})\!)$ on $C^{\infty}(C_{l})\!\subset\! L^{2}(C_{l})$ via $\cD'(C_{l})\hat{\otimes}\ \!C^{\infty}(C_{l})\!)$ is induced from the integrated left regular representation \mbox{$\rho_{\uL}:\cD'(G)\hat{\otimes}\ \!C^{\infty}(G))\rightarrow L(C^{\infty}(G))$:}
\begin{align}
\label{eq:loopphasespaceleftintegratedaction}
\left(\rho^{l}_{\uL}(F)\Psi\right)\left(\{g_{e}\}_{e\in E(\gamma)}\right) & = \int_{G^{\times|E(\gamma)|}}\bigg(\prod_{e\in E(\gamma)}dh_{e}\bigg)F\left(\{(h_{e},g_{e})\}_{e\in E(\gamma)}\right)\Psi\left(\{h^{-1}_{e}g_{e}\}_{e\in E(\gamma)}\right)
\end{align}
for all $F\in\cD'(C_{l})\hat{\otimes}\ \!C^{\infty}(C_{l}),\ \Psi\in C^{\infty}(C_{l})$. Moreover, the action is covariant w.r.t. to the families of maps $\{p^{\uR,\uL*}_{ll'},\alpha^{\uR,\uL}_{l'l}\}_{l \leq l'}$, $\{p^{\uL*}_{ll'},\alpha^{\uL}_{l'l}\}_{l \lessdot_{\uL} l'}$ and $\{p^{\uR*}_{ll'},\alpha^{\uR}_{l'l}\}_{l \lessdot_{\uR} l'}$ (all (sub)diagrams commute):
\begin{align}
\label{eq:loopphasespaceactioncovariance}
\xymatrix@=0.4cm{
C^{\infty}(\Gamma_{l})\!\supset\!\hat{\mathcal{E}}'_{U_{l}}(\mathfrak{c}^{*}_{l})\hat{\otimes}\ \!C^{\infty}(C_{l}) \ar[d]_{\tilde{p}^{\uR,\uL*}_{ll'}} \ar@/_6pc/[dd]_(0.3){\tilde{p}^{\uR,\uL*}_{ll''}\!\!\!\!\!} \ar[rr]^{\hspace{1cm}F^{(W),\varepsilon}} \ar@/^2pc/[rrr]^{Q^{(W)}_{\varepsilon}} &  & \cD'(C_{l})\hat{\otimes}\ \!C^{\infty}(C_{l}) \ar[d]^{\alpha^{\uR,\uL}_{l'l}} \ar@/^4pc/[dd]^(0.3){\!\!\alpha^{\uR,\uL}_{l''l}}|\hole \ar[r]^{\hspace{-1cm}\rho^{l}_{\uL}} & L(C^{\infty}(C_{l}))\circlearrowright C^{\infty}(C_{l})\!\subset\!\! L^{2}(C_{l}) \ar[d]^{p^{*}_{ll'}} \ar@/^6pc/[dd]^(0.3){\!p^{*}_{ll''}} \\
C^{\infty}(\Gamma_{l'})\!\supset\!\hat{\mathcal{E}}'_{U_{l'}}(\mathfrak{c}^{*}_{l'})\hat{\otimes}\ \!C^{\infty}(C_{l'}) \ar[d]_{\tilde{p}^{\uR,\uL*}_{l'l''}} \ar[rr]^{\hspace{1cm}F^{(W),\varepsilon}} \ar@*{[|(6)][white]}@/_0.5pc/[rrr]\ar@/_0.5pc/[rrr]_(0.4){Q^{(W)}_{\varepsilon}} &  & \cD'(C_{l'})\hat{\otimes}\ \!C^{\infty}(C_{l'}) \ar[d]^{\alpha^{\uR,\uL}_{l''l'}} \ar[r]^{\hspace{-1cm}\rho^{l'}_{\uL}} & L(C^{\infty}(C_{l'}))\circlearrowright C^{\infty}(C_{l'})\!\subset\!\! L^{2}(C_{l'}) \ar[d]^{p^{*}_{l'l''}} \\
C^{\infty}(\Gamma_{l''})\!\supset\!\hat{\mathcal{E}}'_{U_{l''}}(\mathfrak{c}^{*}_{l''})\hat{\otimes}\ \!C^{\infty}(C_{l''}) \ar[rr]_{\hspace{1cm}F^{(W),\varepsilon}} \ar@/_2pc/[rrr]_{Q^{(W)}_{\varepsilon}} &  & \cD'(C_{l''})\hat{\otimes}\ \!C^{\infty}(C_{l''}) \ar[r]_{\hspace{-1cm}\rho^{l''}_{\uL}} & L(C^{\infty}(C_{l''}))\circlearrowright C^{\infty}(C_{l''})\!\subset\!\! L^{2}(C_{l''}\!)
}
\end{align}
Consistency w.r.t. to edge inversion and the partial order $\lesssim$ is so far only achieved in the following sense (all (sub)diagrams commute, cp. \eqref{eq:edgeinversionautcompatibilitydiagram}, we abuse notation and use the same notation for the *-morphisms between operators spaces $L(C^{\infty}(C_{l}))$ as for those between kernel spaces $\cD'(C_{l})\hat{\otimes}\!\ C^{\infty}(C_{l})$, $l\in\cL$):
\begin{align}
\label{eq:loopphasespaceactioncovariancerefined}
\xymatrix@=0.2cm{
C^{\infty}(C_{l})\!\circlearrowleft\! L(C^{\infty}(C_{l})\!) \ar@<-6ex>[d]_{p^{*}_{ll'}} \ar@<4ex>[d]^{\alpha^{\uR,\uL}_{l'l}} \ar@<2.5ex>@/^2pc/[rrr]^{p^{*}_{ll^{-1}}} \ar@<1ex>@/^2pc/[rrr]_{\alpha_{l^{-1}l}} & \ar[l]_(0.45){Q^{(W)}_{\varepsilon}} \hat{\mathcal{E}}'_{U_{l}}\!(\mathfrak{c}^{*}_{l})\hat{\otimes}\!\ \!C^{\infty}(C_{l}\!) \ar[r]^(0.45){\tilde{p}^{*}_{ll^{-1}}} \ar[d]_{\tilde{p}^{\uR,\uL*}_{ll'}} & \hat{\mathcal{E}}'_{U_{l^{-1}}}\!(\mathfrak{c}^{*}_{l^{-1}}\!)\hat{\otimes}\!\ \!C^{\infty}(C_{l^{-1}}\!) \ar[r]^{Q^{(W)}_{\varepsilon}} \ar[d]^{\tilde{p}^{\uL,\uR*}_{l^{-1}l'^{-1}}} & L(C^{\infty}(C_{l^{-1}}\!)\!)\!\circlearrowright\! C^{\infty}(C_{l^{-1}}\!) \ar@<-4ex>[d]_{\alpha^{\uL,\uR}_{l'^{-1}l^{-1}}} \ar@<4ex>[d]^{p^{*}_{l^{-1}l'^{-1}}} \\
C^{\infty}(C_{l'})\!\circlearrowleft\! L(C^{\infty}(C_{l'})\!) \ar@<-2.5ex>@/_2pc/[rrr]_{p^{*}_{l'l'^{-1}}} \ar@<-1ex>@/_2pc/[rrr]^{\alpha_{l'^{-1}l'}} & \ar[l]^(0.45){\text{\raisebox{-0.3cm}{$Q^{(W)}_{\varepsilon}$}}} \hat{\mathcal{E}}'_{U_{l'}}\!(\mathfrak{c}^{*}_{l'})\hat{\otimes}\!\ \!C^{\infty}(C_{l'}\!) \ar[r]_(0.5){\text{\raisebox{-0.3cm}{$\tilde{p}^{*}_{l'l'^{-1}}$}}} & \hat{\mathcal{E}}'_{U_{l'^{-1}}}\!(\mathfrak{c}^{*}_{l'^{-1}}\!)\hat{\otimes}\!\ \!C^{\infty}(C_{l'^{-1}}\!) \ar[r]_(0.5){\text{\raisebox{-0.3cm}{$Q^{(W)}_{\varepsilon}$}}} & L(C^{\infty}(C_{l'^{-1}}\!)\!)\!\circlearrowright\! C^{\infty}(C_{l'^{-1}}\!)
}
\end{align}
\begin{Proof}
The statements follow from theorems \ref{thm:loopphasespacepartialordercompatible} and \ref{thm:loopphasespacequant}, and because the $\tilde{p}^{(\uR,\uL)}_{ll'}$ are lifts of the $p_{ll'}$.
\end{Proof}
\end{Corollary}
\subsection{Inductive limit and non-commutative phase spaces}
\label{subsec:indlim}
This subsection is devoted to the question whether it is possible to construct inductive limits of $C^{*}$-algebras from the $\fA_{l},\ l\in\cL,$ which serve as ``non-commutative topological phase spaces'' underlying loop quantum gravity. \\[0.1cm] 
At the level of operators on $C^{\infty}(G)$ via $\rho_{\uL}$, the *-morphisms $\alpha^{\uR,\uL},\gamma,\eta:L(C^{\infty}(G))\rightarrow L(C^{\infty}(G^{\times2}))$ are explicitly given by:
\begin{align}
\label{eq:autopimplementations}
\rho_{\uL}(\alpha^{\uR}(F)) & = U_{\alpha^{\uR}}(\mathds{1}\otimes\rho_{\uL}(F))U^{*}_{\alpha^{\uR}}, & (U_{\alpha^{\uR}}\Psi)(g_{2},g_{1}) & := \Psi(g_{2},g_{2}g_{1}),\ \ \ \Psi\in L^{2}(G^{\times2}), \\ \nonumber
\rho_{\uL}(\alpha^{\uL}(F)) & = U_{\alpha^{\uL}}(\rho_{\uL}(F)\otimes\mathds{1})U^{*}_{\alpha^{\uL}}, & (U_{\alpha^{\uL}}\Psi)(g_{2},g_{1}) & := \Psi(g_{2}g_{1},g_{1}),\ \ \ \Psi\in L^{2}(G^{\times2}), \\ \nonumber
\rho_{\uL}(\gamma(F)) & = U_{\iota}\rho_{L}(F)U^{*}_{\iota}, & (U_{\iota}\Psi)(g) & := \Psi(g^{-1}),\ \ \ \Psi\in L^{2}(G), \\ \nonumber
\rho_{\uL}(\eta(F)) & = \rho_{\uL}(F)\otimes\mathds{1}, & U_{\alpha^{\uR,\uL}}\in U\mathcal{B}(L^{2}(G^{\times2})),\ & U_{\iota}\in U\mathcal{B}(L^{2}(G)).
\end{align}
In this sense $\alpha^{\uR}$ and $\alpha^{\uL}$  are ``twisted'' versions (by the left respectively right action) of the embedding on the second respectively first tensor factor.\\[0.1cm]
It is interesting to note that these maps, apart from $\gamma$, cannot be defined at the level of transformation group $C^{*}$-algebras $C(G)\rtimes_{\uL}G$ and $C(G)\rtimes_{\uL}(G)^{\otimes 2}$, which feature in proposition \ref{prop:loopphasespaceedgeinversion} \mbox{(${\mathfrak{A}}_{l},\ l\in\cL$),} because these algebras are not unital. More precisely, for $A\in\mathcal{K}(L^{2}(G))$, operators of the form $\mathds{1}\otimes A$ or $A\otimes\mathds{1}$ are not compact, and therefore not in $\mathcal{K}(L^{2}(G^{\times2}))$. Thus, if we intend to define a directed system of $C^{*}$-algebras $(\{\fC_{l}\}_{l\in\cL},\{\alpha^{\uR,\uL}_{l'l}\})$ as non-commutative analogue of $|\Lambda|^{1}T^{*}\mathcal{A}$, we need to extend the algebras $\fA_{l}$ to make sense out of \eqref{eq:autopimplementations}.\\
One way to achieve this, which is inspired by the compactification of $\cA$ to $\overline{\cA}$, is to choose \textit{unitisations}\index{algebra!unitisation|textbf} $i_{l}:\fA_{l}\hookrightarrow\fC_{l},\ l\in\cL$ (corresponding to compactifications of the state spaces $\fS_{l}$ of the $\fA_{l}$), i.e. embeddings of $\fA_{l}$ into unital $C^{*}$-algebras $\fC_{l}$ s.t. $i_{l}(\fA_{l})$ is an essential ideal in $\fC_{l}$ (cf. \cite{RaeburnMoritaEquivalenceAnd}). At this point it is not clear which unitisations should be chosen, although it is easy to see that the minimal unitisations $\fA^{\mathds{1}}_{l}\cong\C\mathds{1}+\mathcal{K}(L^{2}(C_{l}))$ via adjoining identites are not sufficient, because $\mathds{1}\otimes A$ and $A\otimes\mathds{1}$ are not in $\C\mathds{1}+\mathcal{K}(L^{2}(G^{\times2}))$ for $A\in\mathcal{K}(L^{2}(G))$. Therefore, we stick to the unique maximal unitisations $M(\fA_{l})\cong\mathcal{B}(L^{2}(C_{l}))$, the \textit{multiplier algebras} of $\fA_{l}$. The latter can be defined as the $C^{*}$-algebras of adjointable operators $\mathcal{B}_{\textup{ad}}(\fA_{l})$ on $\fA_{l}$ as a (left) Hilbert module over itself. The unitisations are then the embeddings $i^{\uL}_{l}:\fA_{l}\hookrightarrow\mathcal{B}_{\textup{ad}}(\fA_{l}),\ i^{\uL}_{l}(a)b=ab,\ a,b\in\fA_{l},$ via (left) multiplication. $M(\fA_{l})$ enjoys several useful properties:
\begin{itemize}
	\item[1.] (unique extension of morphisms): For every $C^{*}$-algebra $\fB$, $\fX$ Hilbert $\fB$-module and non-degenerate *-morphism $\pi_{l}:\fA_{l}\rightarrow\mathcal{B}_{\textup{ad}}(\fX)$, there exists a unique extension $\bar{\pi}_{l}:M(\fA_{l})\rightarrow\mathcal{B}_{\textup{ad}}(\fX)$ s.t. $\bar{\pi}_{l}\circ i^{\uL}_{l} = \pi_{l}$. Moreover, if $\pi_{l}$ is faithful (surjective), so is $\bar{\pi}_{l}$, because $i^{\uL}_{l}(\fA_{l})\subset M(\fA_{l})$ is essential ($\fA_{l}$ is $\sigma$-unital, cf. \cite{RaeburnMoritaEquivalenceAnd}).
	\item[2.] (embedding of $C_{l}$ and $C(C_{l})$): The building blocks $C_{l}\cong G^{\times|E(\gamma)|}$ and $C(C_{l})$ are embedded in $M(\fA_{l})$ in the following sense (a similar statement for $\fA_{l}$ is not true, cf. \cite{WilliamsCrossedProductsOf}): There exist a non-degenerate, faithful *-morphism $i_{C(C_{l})}:C(C_{l})\hookrightarrow M(\fA_{l}$ and a strictly continuous, injective morphism $i_{C_{l}}:C_{l}\hookrightarrow UM(\fA_{l})$ (taking values in unitaries), s.t.
	\begin{itemize}
		\item[(a)] $\forall F\in C(C_{l},C(C_{l})),\ g,h\in C_{l}$ and $f\in C(C_{l})$:
		\begin{align}
		\label{eq:multiplierembedding}
		(i_{C(C_{l})}(f)F)(h) & = fF(h), & , (i_{C_{l}}(g)F)(h) & = \alpha_{\uL}(g)(F(g^{-1}h)).
		\end{align} 
		\item[(b)] The maps are covariant, i.e.
		\begin{align}
		\label{eq:multipliercovariance}
		i_{C(C_{l})}(\alpha_{\uL}(g)(f)) & = i_{C_{l}}(g)i_{C(C_{l})}(f)i_{C_{l}}(g)^{*}.
		\end{align}
		\item[(c)] For every non-degenerate, covariant representation $(\pi,U)$ of $(C(C_{l}),C_{l},\alpha_{\uL})$, the unique extension $\bar{\rho}$ of its integrated form $\rho$ satisfies:
		\begin{align}
		\label{eq:multiplierextension}
		\bar{\rho}(i_{C(C_{l})}(f)) & = \pi(f), & \bar{\rho}(i_{C_{l}}(g)) & = U_{g}.
		\end{align}
	\end{itemize}
	\item[3.] (recovery of $\fS_{l}$): The state space $\fS_{l}\subset\fA^{*}_{l}$ can be recovered from $M(\fA_{l})$ in the following way. Since $\fA_{l}\cong\mathcal{K}(L^{2}(C_{l}))$, we have have $\fA^{*}_{l}\cong\mathcal{S}_{1}(L^{2}(C_{l}))$ (the trace class operators on $L^{2}(C_{l})$). On the other hand, $M(\fA_{l})\cong\mathcal{B}(L^{2}(C_{l}))$, and therefore $M(\fA_{l})$ inherits the structure of a von Neumann algebra with predual $M(\fA_{l})_{*}\cong\mathcal{S}_{1}(L^{2}(C_{l}))$. Now, $\fS_{l}$ corresponds to the set of positive, normalised elements of $\mathcal{S}_{1}(L^{2}(C_{l}))$ (density matrices), which are equivalently characterised as the normal or $\sigma$-weakly continuous states on $\mathcal{B}(L^{2}(C_{l}))$. But, $\sigma$-weakly continuous functionals are the same as $\sigma$-strongly* continuous functionals on $\mathcal{B}(L^{2}(C_{l}))$ (cf. \cite{BratteliOperatorAlgebrasAnd1}), and the $\sigma$-strong* topology on $\mathcal{B}(L^{2}(C_{l}))$ coincides with the strict topology coming from $M(\fA_{l})\cong\mathcal{B}(L^{2}(C_{l}))$. Thus, $\fS_{l}$ can be characterised as the set of \textit{strictly continuous states} on $M(\fA_{l})$. In general, the norm dual of a $C^{*}$-algebra $\fB$ is isometrically isomorphic to the strict dual of $M(\fB)$ with the strong topology (cf. \cite{TaylorTheStrictTopology}).
\end{itemize}
Using property 1, we see that the maps \eqref{eq:autopimplementations} can be defined for $M(C(G)\rtimes_{\uL}G)$ and $M(C(G^{\times2})\rtimes_{\uL}(G^{\times2}))$, if we replace $\rho_{\uL}$ by its unique extension $\bar{\rho}_{\uL}$. Actually, $\bar{\rho}_{\uL}$ provides the natural isomorphism $M(C(G)\rtimes_{\uL}G)\cong\mathcal{B}(L^{2}(G))$. Unfortunately, there is also a drawback in passing from $\fA_{l}$ to $M(\fA_{l})$. Namely, while all the $\fA_{l}$'s are nuclear, and thus are very well-behaved as $C^{*}$-algebras (an inductive limit of such algebras would preserve this porperty), the $M(\fA_{l})$'s are generically not (unless $G$ is finite), as these are type $I_{|G|}$ factors ($|G|$ is the number of elements in $G$, with $|G|=\infty$ for non-finite groups). Therefore, in the present case, it might be advantageous to consider the $M(\fA_{l})$'s as von Neumann algebras, and thus as ``non-commutative measure spaces'' in the usual philosophy. This is further supported by the observation
\begin{align}
\label{eq:multiplierspatialtensorproduct}
M(\fA_{l}\otimes\fA_{l'}) & \cong M(\mathcal{K}(L^{2}(C_{l}))\otimes\mathcal{K}(L^{2}(C_{l'}))) \cong M(\mathcal{K}(L^{2}(C_{l})\otimes L^{2}(C_{l'}))) \\ \nonumber
 & \cong\mathcal{B}(L^{2}(C_{l})\otimes L^{2}(C_{l'}))\cong\mathcal{B}(L^{2}(C_{l}))\bar{\otimes}\mathcal{B}(L^{2}(C_{l'})) \\ \nonumber
 & \cong M(\mathcal{K}(L^{2}(C_{l})))\bar{\otimes}M(\mathcal{K}(L^{2}(C_{l'}))) \\ \nonumber
 & \cong M(\fA_{l})\bar{\otimes}M(\fA_{l'}),
\end{align}
where $\bar{\otimes}$ is the (spatial) tensor product of von Neumann algebras.
\begin{Commentary}
\label{com:compinv}
Before we proceed with the definition of the non-commutative analogue of $|\Lambda|^{1}T^{*}\mathcal{A}$, we add a further comment concerning the partial orders $\leq$ and $\lesssim$. We have noted after equation \eqref{eq:inductivelimorientedconnections}, that the partial order $\leq$ is not directed as opposed to $\lesssim$, $\lessdot_{\uL}$ and $\lessdot_{\uR}$.\\
Therefore, the existence of inductive limits w.r.t. $(\cL,\leq)$ is not ensured, and we would like to pass to limits w.r.t. $(\cL,\lesssim)$. Unfortunately, in view of the non-trivial compatibility conditions between edge composition and inversion, as depicted by the diagrams \eqref{eq:edgeinversioncompatibilitydiagram}, \eqref{eq:edgeinversionautcompatibilitydiagram} and \eqref{eq:loopphasespaceactioncovariancerefined}, it not obvious that this is possible. Nevertheless, we will assume in remainder of this remark, that compatible collections of maps $\{\tilde{p}^{*}_{ll'}\}_{l\lesssim l'}$ and $\{\alpha_{l'l}\}_{l\lesssim l'}$ exist. We will further comment on this issue in the outlook \ref{sec:con}.\\
Alternatively, we may define inductive limits w.r.t. the partial orders $\lessdot_{\uL}$ and $\lessdot_{\uR}$ (cf. \cite{LaneryProjectiveLoopQuantum}). Clearly, $\lessdot_{\uL}$ and $\lessdot_{\uR}$ do not require the implementation of edge inversions as isomorphisms on the level of phase spaces $\Gamma_{l}$ as well as algebras $\fA_{l}\ (or\ M(\fA_{l}))$, because two structured graphs $l,l'$ are only considered to be equivalent in case the underlying oriented graphs are equal $\gamma=\gamma'$.
\end{Commentary}
In summary, we obtain directed systems of (unital) $C^{*}$-algebras $(\{M(\fA_{l})\}_{l\in\cL},\{\alpha_{l'l}\}_{l\lesssim l'})$, \\ $(\{M(\fA_{l})\}_{l\in\cL},\{\alpha^{\uR}_{l'l}\}_{l\lessdot_{\uR} l'})$ and $(\{M(\fA_{l})\}_{l\in\cL},\{\alpha^{\uL}_{l'l}\}_{l\lessdot_{\uL} l'})$, which define unique (unital) inductive limit $C^{*}$-algebras (cf. \cite{KadisonFundamentalsOfThe2}, Section 11.4.)
\begin{align}
\label{eq:inductivelimitcstar}
C^{*}\!\!-\!\!\!\!\varinjlim_{l\in(\cL,\lesssim)}M(\fA_{l}) & = \fA_{\alpha}, & C^{*}\!\!-\!\!\!\!\!\!\!\!\varinjlim_{l\in(\cL,\lessdot_{\uR,\uL})}M(\fA_{l}) & = \fA_{\alpha^{\uR,\uL}}.
 \end{align}
Since we have compatible directed systems of faithful representations $(\{(\bar{\rho}^{l}_{\uL},L^{2}(C_{l}))\}_{l\in\cL},\{U_{l'l}\}_{l\leq l'})$, $(\{(\bar{\rho}^{l}_{\uL},L^{2}(C_{l}))\}_{l\in\cL},\{U_{l'l}\}_{l\lessdot_{\uR} l'})$ and $(\{(\bar{\rho}^{l}_{\uL},L^{2}(C_{l}))\}_{l\in\cL},\{U_{l'l}\}_{l\lessdot_{\uL} l'})$, we have unique inductive limit representations
\begin{align}
\label{eq:inductivelimitrep}
\varinjlim_{l\in(\cL,\lesssim)}(\bar{\rho}^{l}_{\uL},L^{2}(C_{l})) & = (\bar{\rho}_{\uL}, L^{2}(\overline{\mathcal{A}})), & \varinjlim_{l\in(\cL,\lessdot_{\uR,\uL})}(\bar{\rho}^{l}_{\uL},L^{2}(C_{l})) & = (\bar{\rho}_{\uL}, L^{2}(\overline{\mathcal{A}})).
\end{align}
The isometries $U_{l'l}:L^{2}(C_{l})\rightarrow L^{2}(C_{l'})$ are those induced from the maps $p^{*}_{ll'}:C^{\infty}(C_{l})\rightarrow C^{\infty}(C_{l'})$.\\
$\fA_{\alpha^{(\uR,\uL)}}$ could be interpreted as the ``non-commutative topological phase space'' of loop quantum gravity (or a ``non-commutative measure space'' associated with it, if we take the von Neumann algebra point of view).\\
It can be characterised as the $C^{*}$-algebra such that for each $l\in\cL$ there exists an injective \mbox{*-morphism} $\alpha^{(\uR,\uL)}_{l}:\fA_{l}\rightarrow\fA_{\alpha^{(\uR,\uL)}}$ satisfying $\alpha^{(\uR,\uL)}_{l}=\alpha^{(\uR,\uL)}_{l'}\circ\alpha^{(\uR,\uL)}_{l'l}$ for $l\lesssim l'$ ($l\lessdot_{\uR,\uL}l'$), and $\bigcup_{l\in\cL}\alpha^{(\uR,\uL)}_{l}(\fA_{l})\subset\fA_{\alpha^{(\uR,\uL)}}$ is everywhere dense.\\
We also have directed systems of states spaces $(\{\fS_{l}\})_{l\in\cL},\{\alpha^{*}_{l'l}\}_{l\lesssim l'})$, $(\{\fS_{l}\})_{l\in\cL},\{\alpha^{\uR,\uL *}_{l'l}\}_{l\lessdot_{\uR,\uL} l'})$ and $(\{\fS(M(\fA_{l}))\})_{l\in\cL},\{\alpha^{*}_{l'l}\}_{l\lesssim l'})$, $(\{\fS(M(\fA_{l}))\})_{l\in\cL},\{\alpha^{\uR,\uL *}_{l'l}\}_{l\lessdot_{\uR,\uL} l'})$ by duality, which give rise to projective limit state spaces
\begin{align}
\label{eq:projectivelimitstates}
\varprojlim_{l\in(\cL,\lesssim)}\fS_{l} & = \fS^{N}_{\alpha}, & \varprojlim_{l\in(\cL,\lesssim)}\fS(M(\fA_{l})) & = \fS_{\alpha}, \\[0.25cm]
\varprojlim_{l\in(\cL,\lessdot_{\uR,\uL})}\!\!\!\!\fS_{l} & = \fS^{N}_{\alpha^{\uR,\uL}}, & \varprojlim_{l\in(\cL,\lessdot_{\uR,\uL})}\!\!\!\!\fS(M(\fA_{l})) & = \fS_{\alpha^{\uR,\uL}}.
\end{align}
$\fS_{\alpha^{(\uR,\uL)}}$ is isomorphic (continuously in weak* topology) with the state space of $\fA_{\alpha^{(\uR,\uL)}}$, and $\fS^{N}_{\alpha^{(\uR,\uL)}}$ is basic in the latter (cf. \cite{TakedaInductiveLimitAnd}. This implies that we can find a representation $\pi^{(\uR,\uL)}$ of $\fA_{\alpha^{(\uR,\uL)}}$ s.t. the distinguished states (density matrices) of $\pi^{(\uR,\uL)}(\fA_{\alpha^{(\uR,\uL)}})$ coincide with $\fS^{N}_{\alpha^{(\uR,\uL)}}$. Then, the normal states of the weak closure, $\pi^{(\uR,\uL)}(\fA_{\alpha^{(\uR,\uL)}})''$, of $\pi^{(\uR,\uL)}(\fA_{\alpha^{(\uR,\uL)}})$ are the $\sigma$-weakly continuous extensions of elements in $\fS^{N}_{\alpha^{(\uR,\uL)}}$, and
\begin{align}
\label{eq:inductivelimwstar}
W^{*}\!\!-\!\!\!\!\varinjlim_{l\in(\cL,\lesssim)}M(\fA_{l}) & = \pi(\fA_{\alpha})'', & W^{*}\!\!-\!\!\!\!\!\!\!\!\varinjlim_{l\in(\cL,\lessdot_{\uR,\uL})}M(\fA_{l}) & = \pi^{\uR,\uL}(\fA_{\alpha^{\uR,\uL}})'',
\end{align}
which are unique up to (normal) isomorphisms \cite{TakedaInductiveLimitAnd}.\\[0.1cm]
In view of \eqref{eq:projectivelimitstates}, a state $\omega$ on $\fA_{\alpha^{(\uR,\uL)}}$ can be determined from a collection of states $\{\omega_{l}\}_{l\in\cL},\ \omega_{l}\in\fS_{l}$ or $\fS(M(\fA_{l})),$ subject to the consistency conditions coming from the collection of *-morphisms $\{\alpha_{l'l}\}_{l\lesssim l'}$ (or $\{\alpha^{\uR,\uL}_{l'l}\}_{l\lessdot_{\uR,\uL} l'}$), i.e.$\omega_{l}=\omega\circ\alpha^{(\uR,\uL)}_{l}$, $\omega_{l}=\omega_{l'}\circ\alpha^{(\uR,\uL)}_{l'l}$. \\
That the consistency conditions are non-trivial, is due to the composition maps, $\alpha^{\uR,\uL}$, \eqref{eq:algebraautsleftright} used to built the collection $\{\alpha_{l'l}\}_{l\lesssim l'}$ (or $\{\alpha^{\uR,\uL}_{l'l}\}_{l\lessdot_{\uR,\uL} l'}$), as those make $\fA_{\alpha^{(\uR,\uL)}}$ different from an infinite tensor product $\overline{\bigotimes}_{l\in\cL}M(\fA_{l})$, which would be the result of \eqref{eq:inductivelimitcstar}, if we were to use only the map $\eta$ as a building block. \\
On the infinite tensor product any collection of states $\{\omega_{l}\}_{l\in\cL},\ \omega_{l}\in\fS_{l}$ or $\fS(M(\fA_{l})),$ would give rise to a infinite product state $\bigotimes_{l\in\cL}\omega_{l}$ (cf. \cite{KadisonFundamentalsOfThe2}).\\
It is easy to see that the representations $\rho^{l}_{\uL}:\fA_{l}\rightarrow\mathcal{K}(L^{2}(C_{l}))$ arise from the consistent collection of states
\begin{align}
\label{eq:AILstateslocal}
\omega_{l}(F) & =  \int_{G^{\times2|E(\gamma)|}}\bigg(\prod_{e\in E(\gamma)}dh_{e}dg_{e}\bigg)F\left(\{(h_{e},g_{e})\}_{e\in E(\gamma)}\right),\ \ \ F\in C(C_{l},C(C_{l})).
\end{align}
The corresponding consistent collection of complex regular Borel measures $\{\mu_{l}\}_{l\in\cL}, \mu_{l}\in C(C_{l})^{*}$ obtained from the embeddings \mbox{$i_{C(C_{l})}:C(C_{l})\hookrightarrow M(\fA_{l})$} determines the Ashtekar-Isham-Lewandowski measure.
\subsection{Gauge transformations}
\label{subsec:gauge}
Finally, we analyse the behaviour of gauge transformation w.r.t. the Weyl quantisation and the projective limit structure.\\[0.1cm]  
In lemma \ref{eq:loopphasespacegauge}, we have seen how the functionals \eqref{eq:loopphasespacefunctionals} transform w.r.t. gauge transformations $\lambda\in\cG_{\uP}$. On the truncated phase spaces $\Gamma_{l},\ l\in\cL,$ this action corresponds to an action of $G_{l}:=G^{|V(\gamma)|}$ via the strongly Hamiltonian G-actions \eqref{eq:leftrightGactions}:
\begin{align}
\label{eq:truncatedgauge}
\tilde{\lambda}_{l}: G_{l}\times\Gamma_{l} \rightarrow &\ \Gamma_{l}, \\ \nonumber
(\{g_{v}\}_{v\in V(\gamma)},\{(\theta_{e},g_{e})\}_{e\in E(\gamma)}) \mapsto &\ \tilde{\lambda}_{l}\left(\{g_{v}\}_{v\in V(\gamma)},\{(\theta_{e},g_{e})\}_{e\in E(\gamma)}\right) \\ \nonumber
&\ = \{L^{*}_{g_{e(1)}^{-1}}R^{*}_{g_{e(0)}^{-1}}(\theta_{e},g_{e})\}_{e\in E(\gamma)} \\ \nonumber
&\ =\{(Ad^{*}_{g_{e(1)}}(\theta_{e}),g_{e(1)}g_{e}g_{e(0)}^{-1})\}_{e\in E(\gamma)}.
\end{align}
For $l\lesssim l'$, we define (smooth) projections $\pi_{ll'}:G_{l'}\rightarrow G_{l},\ \pi_{ll'}(\{g_{v'}\}_{v'\in V(\gamma')})=\{g_{v'}\}_{v'=v\in V(\gamma)}$, which are compatible with the projections $\tilde{p}^{\uR,\uL}_{ll'}:\Gamma_{l'}\rightarrow\Gamma_{l}$, because
\begin{align}
\label{eq:truncatedgaugecompatibility}
\xymatrix{
G_{l'}\times\Gamma_{l'} \ar[r]^(0.55){\tilde{\lambda}_{l'}} \ar[d]_{\pi_{ll'}\times \tilde{p}^{\uR,\uL}_{ll'}} & \Gamma_{l'} \ar[d]^{\tilde{p}^{\uR,\uL}_{ll'}} \\
G_{l}\times\Gamma_{l} \ar[r]_(0.55){\tilde{\lambda}_{l}} & \Gamma_{l}
}
\end{align}
are commutative diagrams for $l\lesssim l'$. Furthermore, we have the transitivity property $\pi_{ll''}=\pi_{ll'}\circ\pi_{l'l''}$ for $l\lesssim l'\lesssim l''$, and $\tilde{\lambda}_{l}$ is compatible with the cotangent bundle projection, i.e. it is a lift of the action $\lambda_{l}:G_{l}\times C_{l}\rightarrow C_{l}$.\\[0.1cm]
Again, the action \eqref{eq:truncatedgauge} can be transferred to the ``quantum’’ level, where $G_{l}$ acts in an automorphic fashion:
\begin{align}
\label{eq:automorphicgauge}
\alpha^{\lambda}_{l}:\ G_{l} & \rightarrow\aut(\fA_{l}), & & \alpha^{\lambda}_{l}(\{g_{v}\}_{v\in V(\gamma)})(F)(\{((h_{e},g_{e})\}_{e\in E(\gamma)}) \\ \nonumber
\alpha^{\lambda}_{l}:\ G_{l} & \rightarrow\cD'(C_{l})\hat{\otimes}\!\ C^{\infty}(C_{l}), &  & = F(\{(\alpha_{g^{-1}_{e(1)}}(h_{e}),g^{-1}_{e(1)}g_{e}g_{e(0)})\}_{e\in E(\gamma)})
\end{align}
for $F\in C(C_{l},C(C_{l}))$ or $\cD'(C_{l})\hat{\otimes}\!\ C^{\infty}(C_{l})$. \\
The automorphism group $\alpha^{\lambda}_{l}(G_{l})$, as well as its extension $\bar{\alpha}^{\lambda}_{l}(G_{l})$ to $M(\fA_{l})$, is unitarily implemented in $(\rho^{l}_{\uL},L^{2}(C_{l}))$ respectively $(\bar{\rho}^{l}_{\uL},L^{2}(C_{l}))$:
\begin{align}
\label{eq:unitarygauge}
U^{\lambda}_{l}:\ G_{l} & \rightarrow U\mathcal{B}(L^{2}(C_{l})), & \left(U^{\lambda}_{l}(\{g_{v}\}_{v\in V(\gamma)})\Psi\right)(\{g_{e}\}_{e\in E(\gamma)}) & = \Psi(\{g^{-1}_{e(1)}g_{e}g_{e(0)}\}_{e\in E(\gamma)}),
\end{align}
for $\Psi\in L^{2}(C_{l})$. Since $\bar{\rho}^{l}_{\uL}$ is an isomorphism, the implementation is even inner in $M(\fA_{l})$. \\
Furthermore, the various actions \eqref{eq:truncatedgauge}, \eqref{eq:automorphicgauge} \& \eqref{eq:unitarygauge} are consistent with $(\cL,\leq)$, $(\cL,\lessdot_{\uL})$ and $(\cL,\lessdot_{\uR})$ and quantisation via $F^{(W),\varepsilon}$ (or $Q^{(W)}_{\varepsilon}$). We refrain from displaying the projection between the $G_{l}$'s.
\begin{align}
\label{eq:gaugequantconsistency}
{\tiny
\xymatrix@=0.125cm{
 & \hat{\mathcal{E}}'_{U_{l}}(\mathfrak{c}^{*}_{l})\hat{\otimes}\ \!C^{\infty}(C_{l}) \ar[dd]_(0.66){\tilde{p}^{\uR,\uL*}_{ll'}} \ar@/_3.8pc/[dddd]_(0.42){\tilde{p}^{\uR,\uL*}_{ll''}\!\!} \ar[rr]^(0.5){F^{(W),\varepsilon}} &  & \cD'(C_{l})\hat{\otimes}\ \!C^{\infty}(C_{l})\!\circlearrowright\! C^{\infty}(C_{l}) \ar@<-4ex>[dd]_{\alpha^{\uR,\uL}_{l'l}} \ar@<-2.75ex>@/_3.5pc/[dddd]_(0.35){\alpha^{\uR,\uL}_{l''l}\!\!} \ar[dd]^{p^{*}_{ll'}} \ar@/^2.65pc/[dddd]^(0.25){p^{*}_{ll''}} \\
G_{l}\!\times\!\hat{\mathcal{E}}'_{U_{l}}(\mathfrak{c}^{*}_{l})\hat{\otimes}\ \!C^{\infty}(C_{l}) \ar[ur]^(0.4){\tilde{\lambda}^{*}_{l}} \ar[dd]_{\tilde{p}^{\uR,\uL*}_{ll'}} \ar@/_4pc/[dddd]_(0.25){\tilde{p}^{\uR,\uL*}_{ll''}\!\!\!\!\!\!} \ar[rr]^(0.55){F^{(W),\varepsilon}} &  & G_{l}\!\times\!(\cD'(C_{l})\hat{\otimes}\ \!C^{\infty}(C_{l})\circlearrowright C^{\infty}(C_{l})) \ar@<-4ex>[dd]_(0.66){\alpha^{\uR,\uL}_{l'l}} \ar@<-6.7ex>@/_3.8pc/[dddd]_(0.25){\alpha^{\uR,\uL}_{l''l}\!\!}  \ar[dd]^(0.64){p^{*}_{ll'}} \ar@/^2.4pc/[dddd]^(0.2){p^{*}_{ll''}} \ar[ur]^(0.35){(\alpha^{\lambda}_{l},\lambda^{*}_{l}\!)\!\!} & \\
 & \hat{\mathcal{E}}'_{U_{l'}}(\mathfrak{c}^{*}_{l'})\hat{\otimes}\ \!C^{\infty}(C_{l'}) \ar[dd]_(0.35){\tilde{p}^{\uR,\uL*}_{l'l''}} \ar[rr]^(0.7){F^{(W),\varepsilon}} &  & \cD'(C_{l'})\hat{\otimes}\ \!C^{\infty}(C_{l'})\!\circlearrowright\! C^{\infty}(C_{l'}\!) \ar@<-4ex>[dd]_{\alpha^{\uR,\uL}_{l''l'}} \ar[dd]^{p^{*}_{l'l''}} \\
G_{l'}\!\times\!\hat{\mathcal{E}}'_{U_{l'}}(\mathfrak{c}^{*}_{l'})\hat{\otimes}\ \!C^{\infty}(C_{l'}) \ar[ur]^(0.4){\tilde{\lambda}^{*}_{l'}} \ar[dd]_{\tilde{p}^{\uR,\uL*}_{l'l''}} \ar[rr]^(0.55){F^{(W),\varepsilon}} \ar@*{[|(6)][white]}@/_0.5pc/[rrr] &  & G_{l'}\!\times\!(\cD'(C_{l'})\hat{\otimes}\ \!C^{\infty}(C_{l'})\circlearrowright C^{\infty}(C_{l'}\!)) \ar@<-4ex>[dd]_(0.35){\alpha^{\uR,\uL}_{l''l'}} \ar[dd]^(0.37){p^{*}_{l'l''}} \ar[ur]^(0.35){(\alpha^{\lambda}_{l'}\!,\lambda^{*}_{l'}\!)\!\!\!\!} & \\
 & \hat{\mathcal{E}}'_{U_{l''}}(\mathfrak{c}^{*}_{l''})\hat{\otimes}\ \!C^{\infty}(C_{l''}) \ar[rr]^(0.7){F^{(W),\varepsilon}} &  & \cD'(C_{l''})\hat{\otimes}\ \!C^{\infty}(C_{l''})\! \circlearrowright\! C^{\infty}(C_{l''}\!) \\
G_{l''}\!\times\!\hat{\mathcal{E}}'_{U_{l''}}(\mathfrak{c}^{*}_{l''})\hat{\otimes}\ \!C^{\infty}(C_{l''}) \ar[ur]^(0.4){\tilde{\lambda}^{*}_{l''}} \ar[rr]^(0.55){F^{(W),\varepsilon}} &  & G_{l''}\!\times\!(\cD'(C_{l''})\hat{\otimes}\ \!C^{\infty}(C_{l''}) \circlearrowright C^{\infty}(C_{l''}\!)) \ar[ur]^(0.35){(\alpha^{\lambda}_{l''},\lambda^{*}_{l''}\!)\!\!}&
}
}
\end{align}
\section{Conclusions and perspectives}
\label{sec:con}
The Weyl quantisation for loop quantum gravity-type models, which we have constructed in this work, is in some aspects similar to the coherent state techniques, which are usually employed in the discussion of semi-classical limits of loop quantum gravity, because both are based on the truncated phase space formalism for \cite{ThiemannQuantumSpinDynamics7}. Therefore, both methods are best suited to handle semi-classical limits of graph-preserving operators. Clearly, the computation of coherent state expectation values w.r.t. to coherent states, that are built on a single (structured) graph, as was, for example, done in \cite{ThiemannGaugeFieldTheory1, ThiemannGaugeFieldTheory2, ThiemannGaugeFieldTheory3, ThiemannGaugeFieldTheory4, SahlmannCoherentStatesFor, SahlmannTowardsTheQFT1, SahlmannTowardsTheQFT2, GieselAlgebraicQuantumGravity1, GieselAlgebraicQuantumGravity2, GieselAlgebraicQuantumGravity3}, is only meaningful for graph-preserving operators. This was also nicely pointed out in a review on deparametrising models \cite{GieselScalarMaterialReference}, which require the use of graph-preserving operators. Namely, in the Ashtekar-Isham-Lewandowski representation, diffeomorphism-invariant operators have to be graph-preserving. Thus, if the (active) diffeomorphism invariance of the physical Hamiltonian in deparametrising models is to be retained upon quantisation, it is necessary to implement it in a graph-preserving manner. Although, if the framework of algebraic quantum gravity \cite{GieselAlgebraicQuantumGravity1} and its infinite tensor product representation is used, it is sufficient for an operator to preserve the underlying, infinite abstract graph, but not necessarily all possible subgraphs associated with different sectors of the infinite tensor product. On the other hand, if a quantisation by means of graph-changing operators is to be constructed, as e.g. in \cite{DomagalaGravityQuantizedLoop}, where the (physical) Hamiltonian is implemented via graph-changing operators on certain classes of partially diffeomorphism-invariant states, new tools will be necessary.\\
Nevertheless, some preliminary results on the semi-classical limit of graph-changing operators might be obtained through the use of the proposed Weyl quantisation in the following way:\\[0.1cm]
Let us consider a graph-changing operator $O$ and a finite scale of Hilbert spaces, $L^{2}(\overline{\cA}_{\lesssim l})=\bigcup_{l'\lesssim l}L^{2}(C_{l'})$, which is a subspace of $L^{2}(\overline{A})$. Since we have $L^{2}(C_{l'})\subset L^{2}(C_{l})$ for all $l'\lesssim l$, and thus $L^{2}(\overline{\cA}_{\lesssim l}) = L^{2}(C_{l})$, we will obtain a non-trivial restriction $O_{|L^{2}(\overline{\cA}_{\lesssim l})}$ for a large and refined enough ``cut-off graph'' $l$\footnote{This means, that $l$ should contain as least one pair of subgraphs, which are changed into one another}, which is amenable to Weyl quantisation w.r.t. $\Gamma_{l}$. Therefore, the Weyl quantisation can be used to study the family $\{O_{|L^{2}(\overline{\cA}_{\lesssim l})}\}_{l\in\cL}$ of ``graph cut-off'' operators associated with $O$.\\[0.1cm]
A similar scheme can be applied to the coherent state quantisation based on the Segal-Bargmann-Hall transform (see definition III.38 of our second article\cite{StottmeisterCoherentStatesQuantumII}).\\[0.1cm]
A somewhat different line of thought that could be pursued further concerns the compactness problem, which affects the flexibility of the potential implementation of space-adiabatic perturbation theory in loop quantum gravity-type models. Namely, it would be interesting to find out, whether it is possible to choose modified Ashtekar-Barbero variables for loop quantum gravity that are connected to a nilpotent or, more generally, an exponential Lie group. This would make the exponential map a diffeomorphism and lift the problems related to the discrete nature the space of coadjoint orbits. Clearly, such variables would render the Ashtekar-Isham-Lewandowski representation ill-defined, but the construction of the quantum algebras $\fA_{l}\cong C(C_{l})\rtimes_{\uL}C_{l}$ ($l\in\cL$, a structured graph) would still be possible, and the construction of a suitable new representation could be discussed in terms of the projective limit of state spaces, $\overline{\fS}=\varprojlim_{l\in\cL}\fS_{l}$ (cf. \cite{LaneryProjectiveLoopQuantum} for a similar point of view).
In view of full loop quantum gravity, we have not said much about the problem of recovering the phase space $\Gamma=|\Lambda|^{1}T^{*}\mathcal{A}_{\uP}$ of the continuum theory. We have mainly pointed out that the compatibility of the Weyl quantisation is a minimal requirement to discuss the continuum limit by the techniques presented here. In principle, it should be possible to obtain $\Gamma$ along the lines of \cite{ThiemannQuantumSpinDynamics7}, but the correct interplay of the procedure proposed therein with the methods of space-adiabatic perturbation should be verified. Additionally, it might be necessary to adapt the Weyl quantisation to infinite graphs and the associated infinite tensor product construction \cite{ThiemannGaugeFieldTheory4, SahlmannCoherentStatesFor} to allow for a discussion of infinite volume limits.\\[0.1cm]
At this point, we also want to address the technical issue concerning the construction of the projective limit, $\Gamma=\varprojlim_{l\in\cL}\Gamma_{l}$, over structured graphs $l\in\cL$ in the (truncated) phase space quantisation, as it is of utmost importance to the validity of this approach. We have observed in section \ref{sec:apploopphase} (see e.g. commentary \ref{com:compinv}), that the projective structure on the family of (truncated) phase spaces $\{\Gamma_{l}\}_{l\in\cL}$ is only compatible with the (finer) partial orders $\leq$, $\lessdot_{\uL}$ and $\lessdot_{\uR}$ and not necessarily with the partial order $\lesssim$. But, the relations $\lesssim$, $\lessdot_{\uL}$ and $\lessdot_{\uR}$ are those, that identify the Ashtekar-Isham-Lewandowski Hilbert space, $L^{2}(\overline{\cA})=\varinjlim_{l\in\cL}L^{2}(C_{l})$, as a representation space for the quantum algebra $\overline{\fA}=\varinjlim_{l\in\cL}\fA_{l}$. This, is in compliance with the fact that a generalised connection $\bar{A}\in\overline{\cA}$ is completely determined by its values on an oriented representative of a non-oriented graph class.\\
In the commentary \ref{com:compinv}, we only assumed that its is possible to choose collections of maps, $\{\tilde{p}_{ll'}:\Gamma_{l'}\rightarrow\Gamma_{l}\}_{l\lesssim l'}$ and $\{\alpha_{l'l}:L(C^{\infty}(C_{l}))\rightarrow L(C^{\infty}(C_{l'}))\}_{l\lesssim l'}$, that satisfy all transitivity condition induced by the directed partial order $\lesssim$.
Since the main complication in providing such a choice comes from the non-trivial interaction of composition and inversion of edges (diagrams \eqref{eq:edgeinversioncompatibilitydiagram}, \eqref{eq:edgeinversionautcompatibilitydiagram} and \eqref{eq:loopphasespaceactioncovariancerefined}), it is obvious that any oriented representative $\gamma$, together with its oriented  subgraphs $\gamma'\subset\gamma$, of a given non-oriented graph $|\gamma|$ can be given a consistent (w.r.t. $\leq$) choice of maps, $\{\tilde{p}_{ll'}\}_{l\leq l'}$ and $\{\alpha_{l'l}\}_{l\leq l'}$. But, then its is conceivable that, due to the mutual exchange of left and right composition under edge inversion, it is possible to generate relatively consistent choices of maps w.r.t. $\lesssim$, because any other oriented representative $\tilde{\gamma}$ of $|\gamma|$ can be accessed from $\gamma$ via a finite number of single edge inversions.\\
Clearly, this argument only makes the existence of a $\lesssim$-compatible choice of maps for a single non-oriented graph, and its non-oriented subgraphs, plausible. A statement regarding the set of all graphs appears to be difficult. For example, transfinite induction, which would be available, because $(\cL,\lesssim)$ is well-founded ($l\in\cL$ with $\gamma_{l}=\emptyset$ is a minimal element), is not applicable in this case, as it is not necessarily possible to obtain a consistent choice of maps for a given non-oriented graph from its already consistently labelled non-oriented subgraphs without allowing for a relabelling of the latter.\\
on the one hand, it must be admitted that a reconciliation of the (truncated) phase space approach w.r.t. the partial order $\lesssim$ with the usual treatment in terms of the holonomy-flux algebra is still an open problem, which might require further attention. Clearly, this problem also affects the coherent state formalism, which is also based on the (truncated) phase space quantisation considered here.\\
But, on the other hand, in case we use the relation $\lessdot_{\uL}$ and $\lessdot_{\uR}$, everything works fine, and we obtain quantum algebras $\overline{\fA}_{\uL}$ and $\overline{\fA}_{\uR}$, that act in a well-defined fashion on $L^{2}(\overline{\cA})$. The dichotomy between $\lessdot_{\uL}$ and $\lessdot_{\uR}$ reflects the fact, that we have to choose between the left and the right action of the structure group $G$ on itself. These actions are equal for Abelian groups, but they are for only isomorphic via group inversion for non-Abelian groups, which explains why the projective structures w.r.t. $\lessdot_{\uL}$ and $\lessdot_{\uR}$ are related via edge inversion (cp. \ref{eq:edgeinversioncompatibilitydiagram} \& \ref{eq:edgeinversionautcompatibilitydiagram}).\\[0.1cm]
In respect of the aforesaid, it would be interesting to check, whether the Weyl quantisation proposed in this work is also compatible with the projective family of phase spaces constructed in \cite{LaneryProjectiveLoopQuantum}. At first sight this seems possible, because the (truncated) phase spaces used therein are of the cotangent bundle form, $\Gamma_{\eta}\cong T^{*}G^{n_{\eta}}$ ($\eta$ is some label).\\[0.1cm]
With a Weyl quantisation, which is compatible with the (truncated) phase space approach to loop quantum gravity-type models, at our disposal, it appears to be possible to investigate symmetric observables at the classical and quantum level simultaneously. More precisely, if we realise a symmetry by a subgroup of the spatial diffeomorphisms, which act in a natural way on the truncated phase spaces $\Gamma_{l},\ l\in\cL,$ by permutations on the label set $\cL$ (structured graphs), we will be in a position to talk about symmetric functions in $C^{\infty}(\Gamma_{l})$, and thus in $\Cyl^{\infty}(\overline{\Gamma}):=\varinjlim_{l\in\cL}C^{\infty}(\Gamma_{l})$. Moreover, we expect the Weyl quantisation to be covariant w.r.t. to the action of the spatial diffeomorphisms due to the formula \eqref{eq:loopphasespaceleftintegratedaction}, which would entail the invariance of any operator arising as the quantisation of a symmetric function.\\
But, in view of the solution of the spatial diffeomorphism constraint in loop quantum gravity, which makes use of the distributional dual of the linear span of spin network functions, it might be necessary to adapt the Weyl quantisation to handle a suitable distributional extension of $\Cyl^{\infty}(\overline{\Gamma})$.\\
Insights into this aspects of the Weyl quantisation could shed a light onto the question of how to implement semi-classical techniques in a diffeomorphism invariant setting, as well (see above).\\[0.1cm]
In view of the extraction of quantum field theory on curved spacetimes from loop quantum gravity, we have to face yet another type of difficulty, which originates in the well-know methods used to construct linear quantum field theories. On the one hand, space adiabatic perturbation theory relies on the construction of a bundle of Hilbert spaces, $\fH_{\gamma}\cong\fH_{f}$, of the fast sector over the phase space, $\Gamma$, of the slow variables via the (principal symbol of the) projection onto the adiabatically decoupled subspace, $0\rightarrow\fH_{f}\rightarrow\pi_{0}\fH\rightarrow\Gamma\rightarrow0$. Moreover, it is necessary to require the existence of unitary maps between the fibres, $\fH_{\gamma}$, which is typically obstructed by a version of Haag's theorem, unless we allow for some sort of regularisation (in the toy models without regularisation, every fibre carries a quantum field with a different ``mass''). On the other hand, the usual constructions in loop quantum gravity, which are invoked to quantise gravity-matter systems \cite{ThiemannKinematicalHilbertSpaces, ThiemannQuantumSpinDynamics5} (see also \cite{SahlmannTowardsTheQFT1, SahlmannTowardsTheQFT2}), are not anticipated to have this problem due to a natural regularisation of the matter fields by means of the quantisation scheme employed in the gravitational sector, and the use of irregular representations.\\
Thus, further work needs to be invested to gain a better understanding of how the typically regular representations of quantum field theory on curved spacetimes arise in a semi-classical limit of loop quantum gravity with matter content\cite{StottmeisterTheMicrolocalSpectrum}.\\
Another related problem is the actual construction of quantum field theories on specific curved spacetimes via the space-adiabatic approach to loop quantum gravity with matter. Namely, the derivation of spacetime metrics will only be possible, if we extract effective Hamiltonian equations for the gravitational degrees of freedom that give rise to a correspondence between the slow sector's phase space points and said spacetime metrics. But, effective equations, that are obtained in space-adiabatic perturbation theory via a semi-classical limit (Egorov's hierachy), are tied to almost invariant subspaces, which are constructed from spectral bands of the (principal) Hamiltonian symbol that defines the quantum field theories in the fibres of the adiabatic bundle. The upshot of this is, that the resulting spacetime metrics might have a spectral dependence on the quantum matter fields, so-called \textit{rainbow metrics} \cite{AssanioussiRainbowMetricFrom}. Clearly, a further investigation into this aspect is desirable.\\
But, it should be said, that a direct attempt to tackle this problem is only conceivable in symmetry reduced loop quantum cosmology-type models, at the moment. Nevertheless, it would already be a major achievement to rigorously derive the recently constructed loop quantum cosmology-extension of cosmological perturbation theory \cite{AgulloQuantumGravityExtension, AgulloExtensionOfThe}, which invoke a test field approximation (no back reaction), from a model of quantum field theory on a quantum cosmological spacetime (including back reaction) by the methods of space-adiabatic perturbation theory. Regarding full loop quantum gravity, we expect further progress on the extraction of a continuum phase space to be necessary beforehand (see above).
To this end, it is legitimate to say, that the program of adiabatic perturbation theory will require some (substantial) modifications, if a fully satisfactory derivation of quantum field theory on curved spacetimes inside loop quantum gravity is to be obtained along its lines.
\begin{acknowledgments}
\label{sec:ack}
We thank Suzanne Lan{\'e}ry for sharing her expertise on directed partial orders on families of decorated graphs. AS gratefully acknowledges financial support by the Ev. Studienwerk e.V.. This work was supported in parts by funds from the Friedrich-Alexander-University, in the context of its Emerging Field Initiative, to the Emerging Field Project ``Quantum Geometry’’.
\end{acknowledgments}
\bibliography{boa3.bbl, boa3Notes.bib}
\bibliographystyle{aipnum4-1}
\label{sec:ref}
\end{document}